\documentclass[preprint,12pt]{aastex}

\shorttitle{\emph{Spitzer} observations toward high-mass star
forming regions} \shortauthors{Qiu et al.}
\begin{document}

\title{\emph{Spitzer} IRAC and MIPS Imaging of Clusters and
Outflows in 9 High-mass Star Forming Regions}

\author{Keping Qiu\altaffilmark{1,2}, Qizhou Zhang\altaffilmark{2},
S. Thomas Megeath\altaffilmark{3}, Robert A.
Gutermuth\altaffilmark{2}, Henrik Beuther\altaffilmark{4}, Debra
S. Shepherd\altaffilmark{5}, T. K. Sridharan\altaffilmark{2}, L.
Testi\altaffilmark{6,7}, and C. G. De Pree\altaffilmark{8}}
%\affil{Department of Astronomy, Nanjing University, Nanjing, China
%\\ and Harvard-Smithsonian Center for Astrophysics, Cambridge, MA}
\email{kqiu@cfa.harvard.edu}

%\author{Qizhou Zhang}
%\affil{Harvard-Smithsonian Center for Astrophysics, Cambridge, MA}

%\author{S. Thomas Megeath}
%\affil{Department of Physics and Astronomy, University of Toledo,
%Toledo, OH}

%\author{Robert A. Gutermuth}
%\affil{Harvard-Smithsonian Center for Astrophysics, Cambridge, MA}

%\author{Henrik Beuther}
%\affil{Max-Planck-Institute for Astronomy, Heidelberg, Germany}

%\author{Debra S. Shepherd}
%\affil{National Radio Astronomical Observatory, Socorro, NM}

%\author{T. K. Sridharan}
%\affil{Harvard-Smithsonian Center for Astrophysics, Cambridge, MA}

%\author{L. Testi}
%\affil{Osservatorio Astrofisico di Arcetri, Firenze, Italy}

%\and

%\author{C. G. De Pree}
%\affil{Department of Physics and Astronomy, Agnes Scott College,
%Decatur, GA}

\altaffiltext{1}{Department of Astronomy, Nanjing University,
Nanjing, China} \altaffiltext{2}{Harvard-Smithsonian Center for
Astrophysics, Cambridge, MA} \altaffiltext{3}{Department of
Physics and Astronomy, University of Toledo, Toledo, OH}
\altaffiltext{4}{Max-Planck-Institute for Astronomy, Heidelberg,
Germany} \altaffiltext{5}{National Radio Astronomical Observatory,
Socorro, NM} \altaffiltext{6}{Osservatorio Astrofisico di Arcetri,
Firenze, Italy} \altaffiltext{7}{ESO, Karl-Schwarzschild-Str. 2,
D-85748 Garching bei M\"{u}nchen, Germany}
\altaffiltext{8}{Department of Physics and Astronomy, Agnes Scott
College, Decatur, GA}

\begin{abstract}
We present {\it Spitzer} Space Telescope IRAC and MIPS
observations toward a sample of nine high-mass star forming
regions at a distance of around 2 kpc. Based on IRAC and MIPS 24
$\mu$m photometric results and 2MASS \emph{JHKs} data, we carry
out a census of young stellar objects (YSOs) in a $5'\times5'$
field toward each region. Toward seven out of the nine regions, we
detect parsec sized clusters with around 20 YSOs surrounded by a
more extended and sparse distribution of young stars and
protostars. For the other two regions, IRAS 20126+4104 and IRAS
22172+5549, the former has the lowest number of YSOs in the sample
and shows no obvious cluster, and the latter appears to be part of
a larger, potentially more evolved cluster. The deep IRAC imaging
reveals at least twelve outflows in eight out of the nine regions,
with nine outflows prominent in the 4.5 $\mu$m band most probably
attributed to shocked H$_2$ emission, two outflows dominated by
scattered light in the 3.6 and 4.5 $\mu$m bands, and one outflow
standing out from its hydrocarbon emission in the 8.0 $\mu$m band.
In comparison with previous ground-based observations, our IRAC
observations reveal new outflow structures in five regions. The
dramatically different morphologies of detected outflows can be
tentatively interpreted in terms of possible evolution of massive
outflows. The driving sources of these outflows are deeply
embedded in dense dusty cores revealed by previous millimeter
interferometric observations. We detect infrared counterparts of
these dusty cores in the IRAC or MIPS 24 $\mu$m bands. Reflection
nebulae dominated by the emission from UV heated hydrocarbons in
the $8\,\mu$m band can be found in most regions and they may imply
the presence of young B stars.
\end{abstract}


\section{Introduction}
It is well know that high-mass stars form in dense clusters
\citep{Lada03}. The research on embedded young clusters around
young intermediate- to high-mass stars has advanced significantly
via near-infrared surveys \citep[e.g.,][]{Testi99, Gutermuth05,
Kumar06}. The {\it Spitzer} space telescope with its unprecedented
sensitivity at mid-infrared wavelengths provides a new means for
identifying young stellar objects (YSOs) in star forming regions
through detection of dusty disks and envelopes. High sensitivity
in the $3-24$ $\mu$m bands is invaluable for distinguishing young
stars from reddened background stars through infrared excess
arising from circumstellar materials \citep{Allen04, Megeath04,
Whitney04, Gutermuth04, Muzerolle04}. This capability is
particularly valuable for studies of high-mass star forming
regions for two reasons: 1) such regions are typically located
near the Galactic plane where the density of background stars is
high and 2) the member YSOs are often faint due to the large
distances to these regions ($>1$kpc). The faint magnitude of the
YSOs complicates not only the detection of members but also their
identification since the density of background stars rises rapidly
with increasing magnitudes. With the Infrared Array Camera (IRAC)
and Multiband Imaging Photometer for \emph{Spitzer} (MIPS), a
significant census of YSOs can be carried out, which provides a
fossil record of the distribution of star formation sites and
information on the environments in which stars and planets form.
This method is particularly effective for finding extended, lower
density distributions of stars surrounding the dense regions of
embedded clusters.

Extended structures, such as reflection nebulae and outflows, can
also be mapped with the IRAC. Outflows emanating from the central
young stars or protostars are usually detectable in rotational
transitions of CO and other molecules at (sub)millimeter
wavelengths. In IRAC observations, outflows can be revealed from
shocked H$_2$ emission. Hydrodynamic simulations of shock models
predict that H$_2$ emission is particularly strong in the 4.5
$\mu$m band \citep{Smith05}. Since the 4.5 $\mu$m band is
relatively free of the emission from hydrocarbons (probably in the
form of Polycyclic Aromatic Hydrocarbons, or PAHs) compared with
the other three bands, and the 5.8 and 8.0 $\mu$m bands are
significantly less sensitive than the shorter wavelength bands,
composite images with emission from different bands coded in
different colors can be a diagnostic tool for H$_2$ emission in
outflows. A remarkable example of such a detection can be found in
the IRAC observations of the HH 46/47 outflow, a collimated
bipolar outflow emanating from a low-mass protostar, where the
southwestern lobe was clearly seen in the 4.5 $\mu$m band as a
limb-brightened bow-shock cavity \citep{Raga04, Noriega04b}.
\citet{Smith06} reported the IRAC detection of a spectacular
outflow in the high-mass star forming region DR 21 by virtue of
its brightness in the 4.5 $\mu$m band. Further investigations of
the \emph{ISO} SWS spectra of the DR 21 outflow verified that the
H$_2$ lines contribute significantly to all four IRAC bands and
the emission in the 4.5 $\mu$m band is mostly attributed to the
H$_2$ v=0-0 $S(9)$ (4.695 $\mu$m) line. Outflow activities can
also be revealed from scattered light in the IRAC bands. By
comparison of the deep IRAC imaging of L1448 with the radiative
transfer modelling, \citet{Tobin07} demonstrated that the observed
infrared emission in the IRAC bands is consistent with the
scattered light from the central low-mass protostars escaping the
cavities carved by molecular outflows. \citet{Valusamy07}
reprocessed the IRAC observations of the HH 46/47 outflow using a
deconvolution algorithm and found a wide-angle biconical component
in scattered light in addition to the previously reported
bow-shock cavity. Finally, in some cases hydrocarbon emission is
visible from outflow cavities, presumably due to the illumination
of the cavity walls by UV radiation from the central source
\citep{vandenAncker00}.

We have observed a sample of nine high-mass star forming regions
with the IRAC and MIPS instruments (PID: 3528, PI: Qizhou Zhang).
These regions were chosen according to the following criteria: 1)
already imaged with millimeter interferometers; 2) known to have
molecular outflows; 3) outside of the GLIMPSE survey except IRAS
19410+2336; 4) at a distance of $<2.5$kpc ($<5000$AU at the
resolution of IRAC). These regions exhibit far-infrared
luminosities ranging from $10^3$ to $10^5\,L_{\odot}$ (see Table
\ref{source} for source parameters). We carry out a census of YSOs
toward each region, and simultaneously, zoom in on the central
massive star formation sites with the aim of searching for
infrared counterparts of millimeter continuum sources and infrared
emission from outflows. In this paper we present initial results
of the observations. In {\S} 2, we describe the observations. We
present observational results in {\S} 3 and subsequent discussion
in {\S} 4. A brief summary is given in {\S} 5.

\section{Observations and Data Reduction} \label{obs}
The IRAC \citep{Fazio04} observations were obtained from February
to September 2005 except IRAS 19410+2336, which was observed in
October 2004. Each region was observed in the High Dynamic Range
mode with 0.4 second and 10.4 second integration times per dither,
and 16 dithers per map position. The area covered by all four
bands is roughly $5'\times5'$. The total effective integration
time per pixel is 166.4 seconds. The 0.4 second integration frames
were used to obtain photometry for bright sources. The frames were
processed by the Spitzer Science Center (SSC) using the standard
pipeline version S13.2 to produce the standard Basic Calibrated
Data (BCD) products. Additional bright source artifacts
(``jailbar'', ``pulldown'', ``muxbleed'', and ``banding'') were
removed or mitigated using customized IDL scripts developed by the
IRAC instrument team \citep{Hora04, Pipher04}. There are often
``bandwidth'' artifacts in the 5.8 and 8.0 $\mu$m images, where a
decaying trail of the 4th, 8th, and in some cases even 12th pixels
to the right of a bright or saturated spot is found. Currently
there is no correction for this effect. Mosaics were created at
the native instrument resolution of $1.\!''2$ per pixel from the
BCD frames using Gutermuth's WCSmosaic IDL package
\citep{Gutermuth07}. Since the IRAC data is under-sampled and the
point-spread-function (PSF) varies in shape with position and
signal, aperture photometry is currently preferred over the
PSF-fitting photometry. Source finding and aperture photometry
were performed using Gutermuth's IDL photometry and visualization
tool PhotVis, version 1.10 \citep{Gutermuth04, Gutermuth07}. The
radii of the source aperture, the inner and outer boundaries of
the sky annulus were set to $2.\!''4$, $2.\!''4$ and $7.\!''2$,
respectively. IRAC photometry was calibrated using large-aperture
measurements of several standard stars from observations obtained
in flight \citep{Reach05}. There is a $\sim5\%$ calibration
uncertainty due to a position dependent gain for point sources
which is not corrected by the flat field. It is the dominant
source of uncertainty in the photometry of IRAC images. Fluxes at
zero mag are 280.9 Jy at 3.6 $\mu$m, 179.7 Jy at 4.5 $\mu$m, 115.0
Jy at 5.8 $\mu$m, and 64.1 Jy at 8.0 $\mu$m. Magnitudes for 1 DN/s
are set to 19.455 at 3.6 $\mu$m, 18.699 at 4.5 $\mu$m, 16.498 at
5.8 $\mu$m, and 16.892 at 8.0 $\mu$m, using standard aperture
corrections for the radii adopted (see Table 5.7 in IRAC Data
Handbook v3.0). Since regions of active star formation often have
bright nebulosity in the $5.8$ and $8.0\,{\mu}$m bands, we
rejected detections with photometric uncertainties above 0.25 mag
in these two bands, while for the $3.6$ and $4.5\,{\mu}$m bands
detections with uncertainties above 0.2 mag were rejected. The
number of detected sources and the limiting magnitude for each
band toward each region are listed in Table \ref{observation}.

The MIPS \citep{Rieke04} observations were undertaken in the
photometry mode during campaigns separated by less than 1 month
from the IRAC observations with the exception of IRAS 22172+5549
(1.5 months separation) and HH 80-81 (5 months separation). The
target sources were mapped in the 24 and $70\,{\mu}$m bands. The
MIPS 70\,${\mu}$m images were severely saturated and not included
in this paper. The 24\,${\mu}$m observations consist of 400 frames
with an integration of 2.62 seconds per frame, resulting in a
total effective integration time of 1048 seconds. The frames were
processed by the SSC standard pipeline S16.0 for AFGL 5142, IRAS
05358+3543, and G192.16-3.82, and by pipeline S16.1 for the other
regions. Mosaics were created based on the BCD frames using the
SSC MOPEX package, with a pixel size of $1.\!''225$ (half of the
native instrument resolution). The central brightest sources in
all the regions except IRAS 22172+5549 were significantly
saturated. Unsaturated point sources with peaks more than 10 times
the RMS noise level were identified using Photvis 1.10. Aperture
photometry with a 5 pixel aperture and a sky annulus from 12 to 15
pixels was performed on these sources. Multiplying by
$\frac{1}{4}$ to correct for the smaller mosaic pixels and by
1.708 to correct from a 5 pixel aperture to infinite, and using a
zero mag flux of 7.3 Jy, the aperture photometry was converted to
magnitudes \citep{Winston07}. The detections with photometric
uncertainties above 0.25 mag were rejected. The number of detected
sources and the limiting magnitude toward each region are listed
in Table \ref{observation}.

The PSF-fitting photometry is preferred in crowded fields with
multiple point-sources. While in our MIPS 24 $\mu$m images, only
about 10 to 20 unsaturated objects were detected in each field
with a sparse distribution, and in some fields the PSF-fitting
photometry is limited by the lack of ideal PSF-generating stars
which are supposed to be bright, unsaturated, and uncontaminated
by nebulosity. To further evaluate the aperture photometry
quantitatively, we performed PSF-fitting photometry on all the 24
$\mu$m detections using the IDL DAOPHOT package. The reference PSF
was generated from one to two objects chosen from the available in
the field to be most suitable for PSF-generating. We then compared
the returns from the PSF-fitting with the aperture photometry and
found that the majority of unsaturated objects agree with each
other in 0.2 mag; a few objects show differences of 0.3 to 0.4 mag
but these objects are relatively faint and located in a spatially
varying background, thus the differences are probably due to
background subtraction, a limitation to both aperture and
PSF-fitting photometry. Therefore, in Table \ref{photometry} we
show the aperture photometry with the exception of four objects.
These four objects are blended with a neighboring brighter object;
the resulting magnitudes derived from the contaminated aperture
photometry is consequently 0.5 to 0.6 mag brighter than those
determined with the PSF-fitting photometry. We show the
PSF-fitting photometry for these objects and have annotated the
table to identify these objects.

In addition, there are three slightly saturated objects in our
sample (annotated in Table \ref{photometry}); we derived their
magnitudes by manually scaling the PSF, subtracting the PSF from
the image, and comparing the averaged residual of the PSF wing
with a sky value taken from an annulus of 12 to 15 pixels. The
tabulated magnitudes give an averaged residual closest to the sky
value. Since raising/lowering the magnitudes by 0.1 would give an
averaged residual distinctly higher/lower than the sky value, we
adopt an uncertainty of 0.1 mag for these objects.

\section{Results}

\subsection{Young Stellar Objects Census}
In this section, we carry out a census of YSOs and classify them
using their positions in near- to mid-infrared color-color
diagrams. The results of this analysis are shown in Figure
\ref{fig1}, where we display the colors of the detected YSOs in
all nine regions, and Figure \ref{fig2}, where we show
color-composite images of the regions with the identified YSOs
overlaid. We list the number of identified YSOs in Table
\ref{source}, and present the multi-band photometry of each YSO in
Table \ref{photometry}.

Although the criteria we have used to identify YSOs were developed
primarily for low-mass stars, we do not select sources in specific
mass ranges. The census will be dominated by low-mass YSOs, but
intermediate-mass YSOs, such as Herbig Ae/Be stars with disks and
intermediate-mass protostars, may be identified by the same
method. Since we are observing high-mass star forming regions,
high-mass young stars and protostars certainly exist in our
fields. Candidate massive protostellar objects in the IRAC bands
have been identified with the GLIMPSE survey \citep{Kumar07};
however, the colors of those objects are less well understood. It
is unclear whether this method is effective for identifying
high-mass stars with disks or envelopes. In our sample, there are
twelve objects exhibiting obvious infrared excess in one or more
color-color diagrams and also showing evidence of being an
intermediate- to high-mass star (labelled as ``High'' in the last
column of Table \ref{photometry} and marked with blue crosses in
Figure \ref{fig2}). These objects show no or weak dust emission in
the millimeter continuum (except the exciting star of the
H{\scriptsize II} region W75\,N (A), which is within a warm dust
shell) and appear very bright in near-infrared, suggesting the
lack of dense envelopes thus more evolved than typical protostars.
Part of these objects are identified as candidate exciting sources
of the surrounding UV-heated reflection nebulae (\S
\ref{nebulae}). In contrast, the youngest high-mass stars in our
sample are still deeply embedded in dense dusty cores which were
identified as millimeter continuum sources in previous
interferometric observations. Most of these millimeter sources are
not detected shortward of 3.6 $\mu$m, and in some cases even
undetectable at longer wavelengths. Given the extra level of
complication and uncertainty in identifying and classifying these
youngest high-mass objects, we exclude these objects from this
section and discuss them in more detail in \S \ref{msf}.

\subsubsection{IRAC diagram}
\citet{Allen04} and \citet{Megeath04} have established a YSOs
classification scheme in the IRAC color-color diagram [3.6]-[4.5]
vs. [5.8]-[8.0]. This scheme was confirmed by \citet{Hartmann05}
with an IRAC survey of a sample of known Taurus pre-main-sequence
stars, even though some ambiguities were found in a few cases.
Here we adopt this scheme to classify YSOs detected in all four
IRAC bands. The [5.8]-[8.0] color is insensitive to reddening and
is effective at distinguishing between highly reddened background
stars with photospheric emission only and young stars with
circumstellar material. Thus we require $[5.8]-[8.0]>0.2$ for all
the identified YSOs. Bright nebulosity dominated by hydrocarbon
emission features can be found in all the regions and have a very
red [5.8]-[8.0] color.

To assess the impact of nebulosity on creating false detections,
we added artificial stars to each of the mosaics. Each artificial
star was 10 pixel ($12''$) to the west of an actual source that
was detected in the 3.6 and 4.5 $\mu$m bands, and the magnitude of
the artificial star in all four IRAC bands was set to the 3.6
$\mu$m magnitude of the actual source. Thus, the artificial stars
had $[3.6]-[4.5]$ and $[5.8]-[8.0]$ colors equal to $0$. These
sources accurately sampled the range of magnitudes and background
nebulosities found in the image. We then recovered the photometry
to determine how many such stars would be mistaken as having
infrared excesses due to the nebulosity. Although there were about
10 to 30 sources with $[5.8]-[8.0]>0.2$, only 1 to 3 of the
sources had a color of $[3.6]-[4.5]>0.2$.

Hence we require $[3.6]-[4.5]>0.2$ to eliminate pure photospheres
contaminated by hydrocarbon emission nebulosity. Figure
\ref{fig1}a shows the [3.6]-[4.5] vs. [5.8]-[8.0] diagram for
objects merged from all 9 regions and detectable in all four IRAC
bands. In this diagram, objects with $0.2<[3.6]-[4.5]<0.8$ and
$0.4<[5.8]-[8.0]<1.1$ are classified Class II; objects with
$[3.6]-[4.5]>0.8$ and $[5.8]-[8.0]>0.2$, or, $[3.6]-[4.5]\geq0.4$
and $[5.8]-[8.0]>1.1$ are classified protostars (including Class 0
and Class I objects); while objects with $[5.8]-[8.0]>1.1$, which
is consistent with that of protostars, but with
$0.2<[3.6]-[4.5]<0.4$, which is lower than that of protostars, are
classified class I/II. For those detected in the 24 $\mu$m band,
we re-examine them in the IRAC-MIPS diagram.

From both observations and model calculations, the [3.6]-[4.5] vs
[5.8]-[8.0] diagram has been proven quite effective in separating
YSOs with dusty disks or envelops from purely photospheric
emission stars \citep{Hartmann05, Robitaille06}. The principal
ambiguity for the [3.6]-[4.5] vs [5.8]-[8.0] classification scheme
lies in distinguishing protostars from class II objects. From the
direction of the reddening vector in Figure \ref{fig1}a, the
identified protostars immediately above the class II rectangle
($0.8\lesssim[3.6]-[4.5]\lesssim1.2$) are likely reddened class II
objects. In addtion, some of the objects immediately right to the
class II rectangle ($1.1\lesssim[5.8]-[8.0]\lesssim1.5$), which
are identified as protostars or class I/II, are likely class II
objects contaminated by hydrocarbon emission. In addition, some
intrinsic characteristics of YSOs make the classification
ambiguous as well. For example, geometric effects including
bipolar outflow cavities, flattened pattern of envelops and
pole-on geometry will all make the IRAC colors of protostars much
bluer \citep{Hartmann05, Robitaille06}. In a few cases the
discrimination between class II objects and class III/field stars
is also ambiguous; some YSOs with circumstellar disks have an
evacuated or very optically thin inner hole and show little
infrared excess shortward of 8.0 $\mu$m and then can be
misidentified as class III/field stars.

\subsubsection{2MASS-IRAC diagrams}
\begin{figure}
\epsscale{.8} \plotone{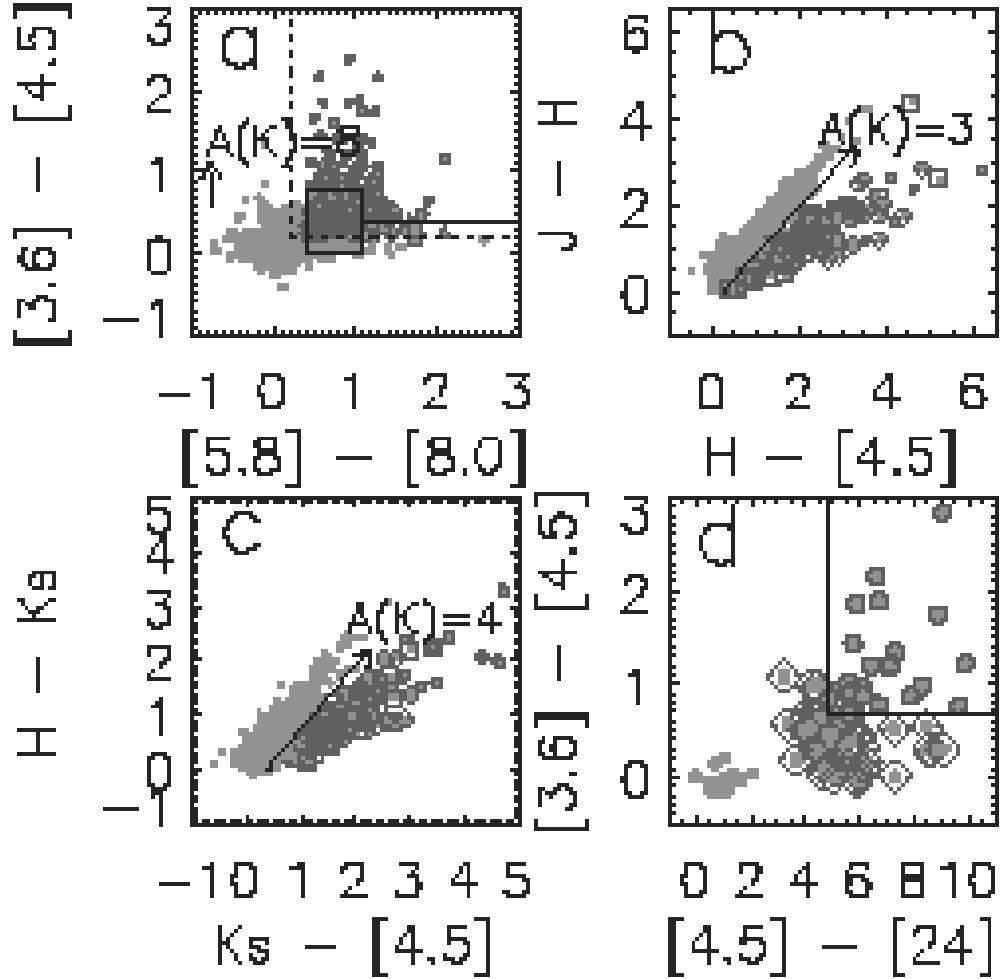} \caption{Color-color diagrams used
to identify and classify YSOs. Protostars, class I/II, and class
II objects are encircled with circles, triangles, and diamonds,
respectively. YSOs that can only be identified in the IRAC-2MASS
diagrams are encircled with squares. Arrows in the first three
diagrams are reddening vectors derived from an averaged reddening
law in nearby star forming regions \citep{Flaherty07}. (a) The
IRAC diagram. The rectangle marks the class II region determined
by \citet{Allen04} and \citet{Megeath04}. The vertical and
horizontal dotted lines represent the requirements of
$[5.8]-[8.0]>0.2$ and $[3.6]-[4.5]>0.2$, respectively. The
horizontal solid line to the right of the class II rectangle
labels the criterion of $[3.6]-[4.5]=0.4$ and $[5.8]-[8.0]>1.1$,
which is used to distinguish between protostars and class I/II
objects \citep{Megeath04}. (b) and (c) IRAC-2MASS diagrams.
Sources more than $1\sigma$ to the right of the reddening vectors
are identified as having intrinsic infrared excesses. (d) The
IRAC-MIPS diagram. The outlined region in the upper right corner
represents the protostar region. An isolated group of sources in
the lower left of the diagram are recognized as class III/field
stars. Objects located outside of the protostar region and not in
the class III/field stars group are classified as class II.
\label{fig1}}
\end{figure}

\begin{figure}
\epsscale{1.} \plotone{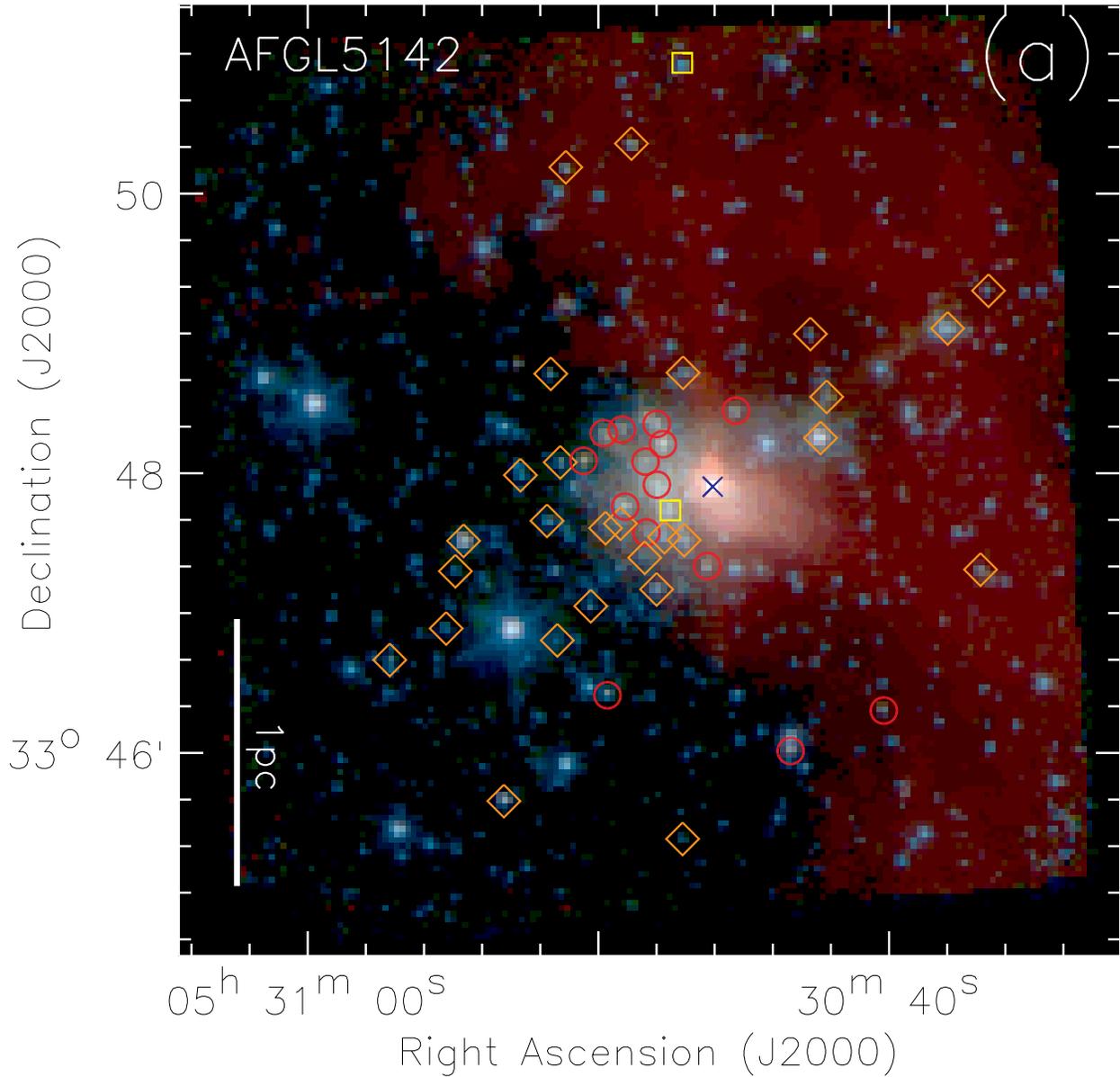} \caption{IRAC 3.6/4.5/8.0 $\mu$m
(Blue/Green/Red) three-color composite images of all 9 regions.
Red circles, red triangles, and brown diamonds represent
protostars, class I/II, and class II objects, respectively. Yellow
squares mark YSOs that can only be identified in the IRAC-2MASS
diagrams. Blue crosses label candidates of intermediate- to
high-mass young stars (see ``high'' sources in Table
\ref{photometry} for details). \label{fig2}}
\end{figure}

The IRAC 5.8 and 8.0 $\mu$m bands are far less sensitive than the
3.6 and 4.5 $\mu$m bands (see the limiting magnitudes in Table
\ref{observation}), and are often dominated by bright nebulosity
in observations of active star forming regions. For this reason,
we combine the IRAC photometric results with data from the 2MASS
\emph{JHKs} Point Source Catalog. We limit the \emph{JHKs}
photometric uncertainty to be less than 0.1 mag. Since the 4.5
$\mu$m band is the most sensitive band to YSOs among all four IRAC
bands \citep{Gutermuth04}, we use the $J-H$ vs. $H-[4.5]$ and
$H-Ks$ vs. $Ks-[4.5]$ diagrams (Figures \ref{fig1}b, \ref{fig1}c)
to identify objects with infrared excess emission but not detected
in the IRAC longer wavelength bands. In Figures \ref{fig1}b and
\ref{fig1}c, objects with the $H-[4.5]$ or $Ks-[4.5]$ colors more
than $1\sigma$ beyond the reddening vectors are identified as
objects with intrinsic infrared excess. The extinction law in the
mid-infrared wavelengths toward star-forming regions, in
particular toward distant high-mass star forming regions, has not
been well established. Most recently, \citet{Flaherty07} derived
the interstellar extinction laws in IRAC bands toward five nearby
star-forming regions. We adopt the average of the extinction laws
derived by Flaherty et al. for our sample. From the dispersion of
the objects left to the reddening vectors in these two diagrams,
it seems the adopted reddening law is applicable for our high-mass
star forming regions. The positions offsets of the reddening
vectors in these two diagrams, which were determined by
\citet{Winston07} for Serpens, also appear to be a good
approximation for our regions. However, the completeness of the
YSOs identification from these two diagrams are significantly
limited by the modest sensitivity of the 2MASS survey. Deeper
near-infrared observations will not only improve the completeness
but also help to derive a more accurate reddening law toward each
region, which would improve the reliability of the YSOs
identification.

\subsubsection{IRAC-MIPS diagram}
Including data at wavelengths longward of 20 $\mu$m is valuable
for more reliable classification of YSOs \citep{Robitaille06}. In
Figure \ref{fig1}d, we use the [3.6]-[4.5] vs. [4.5]-[24] diagram
to classify YSOs which are detectable both in the two most
sensitive IRAC bands and in the MIPS 24 $\mu$m band. Sources with
spectral indexes greater than -0.3 have IRAC-MIPS colors
$[3.6]-[4.5]>0.652$ and $[4.5]-[24]>4.761$. These sources would
include both the flat spectrum and class I objects in the
classification scheme of \citet{Greene94}. In addition, the high
sensitivity of {\it Spitzer} data allows for the detection of
previously identified class 0 sources in mid-infrared
\citep{Noriega04a, Rho06, Hatchell07, Winston07}; consequently,
class 0 sources can also be identified by these criteria. Since an
infalling envelope is required to explain the SEDs of flat
spectrum, class I and class 0 sources \citep{Calvet94, Whitney03},
we refer to the combined set of these sources as protostars. In
Figure \ref{fig1}d, there is clearly an isolated group close to
(0, 0). The objects in this group are class III/field stars whose
emission is dominantly from photospheres. In this diagram objects
outside of the protostar region and not in the class III/field
star group are classified as class II. There is an obvious gap
between protostars/class II objects and classIII/field stars,
suggesting this diagram is very effective in distinguishing YSOs
with circumstellar materials from purely photospheric objects.
From the objects that can be classified both in the [3.6]-[4.5]
vs. [5.8]-[8.0] and in the [3.6]-[4.5] vs. [4.5]-[24] diagrams, we
find that the two classification schemes are in general
consistent, otherwise we adopt the latter if there is an
ambiguity. With the IRAC-MIPS diagram we also identify 15 class II
objects and 2 protostars that cannot be identified in any of the
other three color-color diagrams.

\subsubsection{Removing extragalactic contaminants}
There are mainly two classes of extragalactic contaminants that
can be misidentified as YSOs \citep{Stern05}. One is star forming
galaxies and narrow-line active galactic nuclei (AGN) which have
growing excess at 5.8 and 8.0 $\mu$m due to hydrocarbon emission.
The other is broad-line AGN which have IRAC colors very similar to
that of bona fide YSOs. \citet{Gutermuth07} have developed a
method based on the Bootes Shallow Survey data to substantially
mitigate these contaminants. In this method, hydrocarbon emission
sources, including galaxies and narrow-line AGN, can be eliminated
by their positions in the [4.5]-[5.8] vs. [5.8]-[8.0] and
[3.6]-[5.8] vs. [4.5]-[8.0] diagrams, while broad-line AGN, which
are typically fainter than YSOs in the {\it Spitzer} bands, are
identified by their positions in the [4.5] vs. [4.5]-[8.0]
diagram. Here we adopt this method and lower the [4.5] limit in
the [4.5] vs. [4.5]-[8.0] diagram by 1 mag since our regions are
far more distant. To further remove AGN, we set $[3.6]<16$ for
YSOs not detected in the 8.0 $\mu$m band and require $[24]<8.5$
for those only identifiable in the [3.6]-[4.5] vs. [4.5]-[24]
diagram. We filter out a total of 11 extragalactic contaminants in
all nine regions.

\subsubsection{Completeness of the census}
Although we are undertaking deep observations toward relatively
nearby high-mass star forming regions, the completeness of the
YSOs census is limited by a few factors. Bright extended
nebulosity in the IRAC bands can be found in all the regions,
which significantly limit the point source detection in these
areas. The YSOs identification from the 2MASS-IRAC diagram is
limited by the modest sensitivity of the 2MASS survey, while that
from the IRAC-MIPS diagram suffers from the significant saturation
in the MIPS 24 $\mu$m image caused by the central luminous
sources. To further evaluate the completeness of the census
quantitatively, we plot histograms of the sources which are
counted within 0.5 mag bins of the 3.6 $\mu$m photometry in Figure
\ref{complete}. In this plot, all the 3.6 $\mu$m detections with
signal to noise ratios sufficient to be identified by our point
source filter are plotted with the gray solid lines, and sources
with corresponding multi-band photometry so that they can be
placed on one or more color-color diagrams for an assessment of
infrared excess are plotted with the dark dashed lines. The
identified YSO members are also plotted with the gray dotted
lines. There are two types of completeness in this diagram. The
first is the completeness in the 3.6 $\mu$m band, which declines
strongly where a peak is evident in the solid gray line histogram.
The second is the coupled multi-band completeness to infrared
excess sources, which is approximately given where the dark dashed
line diverges from the solid gray line. Since the 3.6 $\mu$m band
detects sources of fainter magnitudes than the other bands, we
define the completeness magnitude as the point where the
multi-band histogram declines more than 30\% with respect to the
3.6 $\mu$m histogram. A vertical dash-dotted line in each plot
indicates the completeness magnitude. For sources with the 3.6
$\mu$m photometry brighter than the completeness magnitude, we
calculate the ratio of the number of sources that can be placed on
one or more color-color diagrams for the search of infrared excess
to the number of the total detections (Table \ref{fraction}). The
actual completeness to infrared excess sources is presumably
higher since they will have stronger emission in the IRAC bands
than field stars and young stars without disks.

\subsection{Central Massive Star formation Sites} \label{msf}
Previous interferometric studies have identified dusty cores at
the center of each region through the millimeter continuum
emission, and so far five regions show multiple or a cluster of
dusty cores. Of these millimeter sources, the deep IRAC imaging
detects six at 8.0 $\mu$m, five at 5.8 $\mu$m, three at 4.5
$\mu$m, and two at 3.6 $\mu$m. The low detection rate is
consistent with the deep IRAC observations of the other two nearby
(1.7 kpc) high-mass star forming regions, NGC 6334 I and NGC 6334
I(N), where only one out of eleven SMA 1.3 mm sources shows a
detectable infrared counterpart \citep{Hunter06}. \citet{Kumar07}
suggested that the IRAC band emission of candidate massive
protostellar objects arises from the luminous envelopes around the
protostars rather than their photospheres or disks. In our sample,
the number of detected infrared counterparts increases with the
increasing wavelength of the band, despite the fact that the 3.6
and 4.5 $\mu$m bands are much more sensitive than the 5.8 and 8.0
$\mu$m bands. This implies a steeply rising SED in the IRAC bands,
which is consistent with the interpretation that the emission is
mostly coming from the dense envelopes. All the six millimeter
sources detected in one or more IRAC bands appear significantly
saturated in the MIPS 24 $\mu$m band, while there are two
millimeter sources detected at 24 $\mu$m without saturation but
undetectable in any of the four IRAC bands. We list the IRAC or
MIPS 24 $\mu$m photometry of detected millimeter sources in Table
\ref{irmm}. In our sample, high-mass stars deeply embedded in
dusty cores are early-B types estimated from far-infrared
luminosities, ionizing UV radiation fluxes (if UC H{\scriptsize
II} regions detected), and core masses. Also considering they are
associated with molecular outflows indicative of active accretion,
we refer to these deeply embedded B stars as proto-B stars.

In all nine regions, massive molecular outflows have been
identified in millimeter lines of CO and/or SiO. In the IRAC
observations, the central high-mass sources are found to be
associated with extended nebulosity. Several mechanisms produce
extended nebulosity in the IRAC bands. Prominent features from
hydrocarbons appear in the 3.6, 5.8, and 8.0 $\mu$m bands, with
the strongest emission features in the 8.0 $\mu$m band
\citep{Werner04}. Vibrational and rotational H$_2$ lines appear in
all four bands, with the weakest and most spatially confined
emission in the 3.6 $\mu$m band \citep{Smith06, Smith05}; this
H$_2$ emission is most apparent in the 4.5 $\mu$m band due to the
high sensitivity of this band and presence of strong rotational
lines \citep{Smith05}. Finally scattered light may also produce
nebulosity which is expected to be the strongest in the 3.6 $\mu$m
band. To distinguish between these different mechanisms, we employ
3.6/4.5 $\mu$m two-color and 3.6/4.5/8.0 $\mu$m three-color
composite images to highlight emission from shocked H$_2$ in
outflows, scattered light from outflow cavities, and emission from
UV heated hydrocarbons. We identify a total of twelve outflows in
eight out of the nine regions. Reflection nebulae and externally
heated rimmed clouds are detected as well.

In the following subsections, we discuss each region separately.
For Figures \ref{afgl5142} to \ref{w75n}, the 3.6, 4.5 and 8.0
$\mu$m emissions in color composites are coded in blue, green and
red, respectively; millimeter continuum cores from previous
observations are marked as plus signs; CO and SiO outflows from
previous observations are shown in contours; additional symbols
(arrows, dashed curves, etc.) are merely to highlight the IRAC
outflows from the ambient reflection nebulae or scattered light.

\subsubsection{AFGL 5142} \label{msf_afgl5142}
In Figure \ref{afgl5142}a, a short jet in the east and a long jet
in the west are detected. The short jet coincides with the axis of
one of the three jet-like molecular outflows detected in the CO
(2-1) and SO ($6_5-5_4$) with the SMA \citep[outflow B in][see
contours in Figure \ref{afgl5142}b]{Zhang07}. The long jet, on the
other hand, coincides with the axis of an extended remnant outflow
detected in CO (2-1) with the CSO telescope \citep{Hunter95}. The
emission from these two jets is most prominent in the 4.5 $\mu$m
band, in agreement with the previous ground-based 2.12 $\mu$m
H$_2$ v=1-0 $S(1)$ observations \citep{Hunter95, Chen05}.
Apparently the IRAC imaging of this region reveals internal
driving agents of two molecular outflows.

Toward the center of AFGL 5142, \citet{Zhang07} detected five
millimeter continuum sources (MM1 to MM5 in Figure
\ref{5142core}). In Figure \ref{5142core}, a bright source
residing $0.\!''8$ northwest to MM1 (marked with a star symbol) is
detected in the 2MASS \emph{Ks} band, all four IRAC bands and MIPS
24 $\mu$m band (see the detection at
$\alpha($J$2000.0)=05^{\mathrm h}30^{\mathrm m}47.98^{\mathrm s}$,
$\delta($J$2000.0)=33^{\circ}47'54.9''$ in Table
\ref{photometry}). It is classified as a protostar in the
[3.6]-[4.5] vs. [5.8]-[8.0] and [3.6]-[4.5] vs. [4.5]-[24]
diagrams. The projected offset between this source and MM1 is
marginally within the IRAC astrometry allowance. Thus it is still
possible that this source is the infrared counterpart of MM1.
However, \citet{Zhang07} detected a forest of molecular lines
toward MM1 and suggested it to be the driving source of a jet-like
molecular outflow, which implies that the protostar deeply
embedded in MM1 is actively accreting. Such a protostellar core is
usually obscured in $K$ band \citep[e.g.][and observations of
other regions in our sample]{Walther90, Aspin91}. Considering the
source is detected in the 2MASS \emph{Ks} band, a survey with
limited sensitivity, we suggest that it is most likely a different
source rather than the infrared counterpart of MM1. But the 24
$\mu$m emission can still arise from MM1 and/or other millimeter
sources given its relatively low resolution ($\sim5''$), thus we
list the 24 $\mu$m photometry in Table \ref{irmm}. The 24 $\mu$m
detection is located at the bright PSF wing of a nearby saturated
source, resulting in a potential overestimate of the flux by about
$0.3$ to $0.4$ Jy. The driving sources of the multiple CO/SO
outflows have not been unambiguously identified. \citet{Zhang07}
suggested MM3 may be the powering source of the CO outflow
coincident with the short jet in Figure \ref{afgl5142}b. The long
jet is apparently powered by a relatively more evolved young B
star (denoted by a blue cross in Figure \ref{fig2}a).

\subsubsection{IRAS 05358+3543} \label{msf_i05358}
In Figure \ref{i05358}, three highly collimated jets emanating
from a group of millimeter continuum sources are detected in the
central part of IRAS 05358+3543 (hereafter referred as IRAS
05358). Jet1 coincides a well collimated outflow in CO (1-0)
reported by \citet{Beuther02} (see contours in Figure
\ref{i05358}). Jet2 coincides with the axis of a high-velocity CO
outflow (outflow B in \citet{Beuther02}). Several knots from these
two jets are almost exactly coincident with the previously
identified 2.12 $\mu$m H$_2$ knots \citep{Porras00, Jiang01,
Beuther02, Kumar02}. \citet{Beuther02} did not report the
identification of Jet3, but weak emission features coincident with
this jet can be found in their 2.12 $\mu$m H$_2$ image.
\citet{Kumar02} suggested the existence of this jet, but the
emission appears to be weaker since their H$_2$ observations are
less deep than that of \citet{Beuther02}. In the IRAC imaging
(Figure \ref{i05358}), a chain of emission knots in this jet can
be identified and even its tip can be identified despite the
contamination by a nebulous knot. There is no reported detection
of an outflow toward this jet in millimeter CO or SiO
observations. However, in the northern lobe of the SiO (2-1)
outflow proposed by \citet{Beuther02} (their Fig. 8b), the highest
contours show elongation coincident with Jet3. Considering SiO is
a good shock tracer at (sub)millimeter wavelengths, the elongated
SiO emission could be at least partially powered by Jet3.

At the center of this region, \citet{Beuther02, Beuther07}
detected three millimter continuum sources at $\sim2-3''$
resolutions (marked as MM1 to MM3 in Figure \ref{05358core}). The
brightest emission feature in Figure \ref{05358core} coincides
with MM1; two northern ``red'' features (denoted by two white
dashed circles) are not real but ``bandwidth'' artifacts. An
investigation of the emission in individual bands shows that
emission features coincident with MM2 and MM3 are discernible in
the 3.6 and 4.5 $\mu$m bands but is not identified by our point
source filter. The emission feature coincident with MM1 is
marginally discernible in the 3.6\,$\mu$m band, and identified as
a point source longward of 4.5 $\mu$m. The MIPS 24 $\mu$m emission
from this area is significantly saturated. From its extremely red
IRAC colors ([4.5]-[5.8] = 1.33 and [5.8]-[8.0] = 1.15) and
position coincidence with MM1, we suggest this source to be the
infrared counterpart of MM1 and list its photometry in Table
\ref{irmm}. At higher resolutions ($\lesssim1''$), the MM1 core
splits into two sub-cores at 875 $\mu$m and 1.2 mm
\citep{Beuther07}. One of the two sub-cores coincides with a point
source at 7.9 $\mu$m detected with the Gemini North telescope
\citep{Longmore06}. The sub-structure of MM1 cannot be resolved in
the IRAC observations. The flux of 0.586 Jy at 8.0 $\mu$m
($[8.0]=5.098$) is consistent with the flux of 0.68 Jy at 7.9
$\mu$m reported in \citet{Longmore06}, suggesting that the flux of
the IRAC source is dominated by the sub-core coincident with the
7.9 $\mu$m detection. MM1 (or more precisely, one of the two
sub-cores in MM1) is suggested to be the driving source of Jet1
\citep{Beuther07}. A reliable one-to-one association between Jet2,
Jet3 and any two millimeter sources in this region cannot be
established with the current data.

\subsubsection{G192.16-3.82} \label{msf_g192}
\citet{Shepherd98} detected a west-east bipolar outflow in CO
(1-0) in G192.16-3.82 (hereafter referred as G192). The eastern
lobe of this outflow is detected in the 3.6 and 4.5 $\mu$m bands.
In Figure \ref{g192}, the most prominent features of this lobe are
``green'' nebulosities within the area outlined by two dashed
lines. The emission to the west of these features is confused by
the bright reflection nebula, but a ``blue'' V-shaped structure,
with the millimeter source at the tip, is still discernable in the
three-color composite of the field (Figure \ref{fig2}c). The
overall structure of the eastern lobe in the IRAC imaging, i.e.,
the ``green'' features as well as the V-shaped inner part, roughly
coincides with but far exceeds the eastern lobe of the CO outflow.
The prominence in the 4.5 $\mu$m band and coincidence with
previous H$\alpha$ and [S{\scriptsize II}] detections
\citep{Devine99} strongly suggest that the ``green'' nebulosities
are shock excited H$_2$ knots. The base of the western lobe of the
CO outflow is also detected in the IRAC imaging, with the
prominent feature being a candle-flame-shaped structure
(delineated by a dashed curve in Figure \ref{g192}, but more
prominent in Figure \ref{fig2}c); this structure is previously
detected in the $K$-band imaging \citep{Shepherd98, Devine99}. One
interesting aspect of this outflow is that although the eastern
lobe is filled with shocked H$_2$ knots, a collimated component
cannot be found either in our IRAC imaging or previous
observations. The CO outflow appears to be driven by a wide-angle
wind revealed in the IRAC, H$\alpha$ and [S{\scriptsize II}]
observations.

A proto-B star surrounded by an UC H{\scriptsize II} region and
embedded in a dense dusty core was suggested to be the driving
source of the outflow \citep{Shepherd99, Shepherd04a}. A bright
source coincident with the millimeter continuum peak is detected
in the IRAC bands and show very red IRAC colors (Figure
\ref{192core}). We suggest it to be the infrared counterpart of
the millimeter source and list its photometry in Table \ref{irmm}.
A detailed comparison between the infrared and millimeter emission
in the region will be presented in a future work (Shepherd et al.,
in prep.).

\subsubsection{HH 80-81} \label{msf_hh8081}
An extraordinarily well-collimated and powerful radio jet was
previously detected in HH 80-81 \citep[][see contours in Figure
\ref{h8081}]{Marti93,Marti95,Marti98}. Having the radio jet right
at the axis, the {\it Spitzer} data for the first time reveals a
bipolar cone-shaped cavity with its emission most prominent in the
8.0 $\mu$m band with an extent of about 4pc (Figure \ref{h8081}).
Low-mass protostellar outflows are often found to consist of two
components, with an axial narrow jet-like component surrounded by
a wide-angle biconical component \citep{Lee06, Palau06}. It is
unclear whether outflows emanating from high-mass young stars or
protostars may exhibit two components similar to low-mass
outflows. Our IRAC imaging as well as previous radio continuum
observations of the HH 80-81 outflow provides the best case of a
two-component outflow emanating from a $10^4L_{\odot}$ source. The
remarkable prominence in the 8.0 $\mu$m band suggests that the
emission of the IRAC biconical structure is from hydrocarbons
heated by UV photons from the central source. Single-dish
observations revealed a large bipolar outflow in CO (1-0)
\citep{Yamashita89, Benedettini04}, with the northern lobe being
blueshifted and the southern lobe redshifted. The orientation of
the CO outflow is roughly consistent with the IRAC outflow
structure. In the three-color composite image (Figure
\ref{fig2}d), a curved structure in the northern lobe stands out
in the 3.6 $\mu$m emission (blue). This curved structure is very
bright in previous near-infrared $K$-band observations and has
been suggested as the wall of a parabolic cavity \citep{Aspin91}.
The dominant 3.6 $\mu$m emission from this structure may be partly
explained by the scattered light in this structure being enhanced
by forward scattering and lower extinction. The 8.0 $\mu$m
emission within this structure is spatially limited to the inner
edge of the parabolic wall. This may suggest that part of the
northern lobe cavity is shielded from the UV photons by gas and
dust in the outflow and perhaps by the circumstellar environment
around the central source. Two additional reflection nebulae are
found on both sides of the bipolar cavity; these appear to be
heated by two young B stars (see Table \ref{photometry}) and not
directly connected to the outflow cavity.

The central driving source of the outflow has been detected in the
centimeter \citep{Marti93, Marti99} and millimeter \citep{Gomez03}
continuum. Mid-infrared emission from this source has been
detected at 4.7 to 13 $\mu$m in the ground-based observations
\citep{Aspin94, Stecklum97}. In Figure \ref{8081core}, the
centroid of the emission feature in the 3.6 $\mu$m band is about
$2''$ northeast offset from the millimeter peak. This feature
coincides with a $K$-band knot detected in previous ground-based
observations \citep[their IRS 2,][]{Aspin91, Aspin94, Stecklum97}.
From its high polarization and relatively large FWHM,
\citet{Aspin91} argued that it is a nebulous knot of gas/dust
rather than a point source. The emission feature in the 4.5 $\mu$m
band shows elongation toward the position of the millimeter peak.
The centroid of the emission feature in the 5.8 and 8.0 $\mu$m
bands, discarding the ``bandwidth'' artifact in the north,
coincides with the millimeter peak. We suggest that the emission
in the 3.6 and 4.5 $\mu$m bands is mostly attributed to the IRS 2
knot, while that in the long wavelength bands is mostly coming
from the millimeter core and their photometry is listed in Table
\ref{irmm}. The MIPS 24 $\mu$m emission in the central part of
this region is significantly saturated.

\subsubsection{IRAS 19410+2336} \label{msf_i19410}
As outlined with a dashed ellipse in Figure \ref{i19410}, an
elliptical nebula orientated in the northeast-southwest direction
can be identified from the extended emission in the central part
of IRAS 19410+2336 (hereafter IRAS 19410). In interferometric CO
observations, multiple outflows emanating from the center of this
region show very complex morphologies \citep{Beuther03}. The axis
of one outflow proposed by Beuther et al., which is shown in
contours in Figure \ref{i19410}, coincides with the major axis of
the IRAC elliptical nebula. Jet-like 2.12 $\mu$m H$_2$ emission
along the axis of this outflow was detected as well
\citep{Beuther03}. The elliptical nebula, with two local cavities
at the ends of the major axis, is only detectable in the 3.6 and
4.5 $\mu$m bands and appears more prominent in the 3.6 $\mu$m
band, suggesting the emission is dominated by scattered light from
the central source. Apparently the IRAC imaging of the IRAS 19410
outflow reveals a less collimated component than the 2.12 $\mu$m
H$_2$ jet.

The driving source of this outflow is suggested to be deeply
embedded in a dusty core traced by millimeter continuum emission
at a resolution of $5.5''\times3.5''$ \citep{Beuther03}. The
infrared emission from this millimeter source is detected in all
four IRAC bands and the photometry is listed in Table \ref{irmm}.
There is a 2MASS $Ks$ band detection, with the photometry of
$10.802\pm0.044$, at $\sim0.7''$ away from the millimeter
continuum peak. As we discussed in \S \ref{msf_afgl5142}, it is
more likely another star rather than the infrared counterpart of
the millimeter source. The MIPS 24 $\mu$m emission from the source
is significantly saturated. At a higher resolution
($1.5''\times1''$), the millimeter core splits into multiple
sources \citep{Beuther04a}; our IRAC observations cannot resolve
those sources. A detailed comparison between the IRAC observations
and high-spatial-resolution millimeter continuum observations of
this region will be presented in a future work (Rodon et al., in
prep.).

\subsubsection{IRAS 20126+4104} \label{msf_i20126}
In Figures \ref{i20126}a \& \ref{i20126}c, a bipolar outflow in
IRAS 20126+4104 (hereafter referred as IRAS 20126) is detected as
a limb-brightened biconical cavity. The cavity structure, in
particular the wall of the northwestern lobe, is clearly detected
in the 3.6, 4.5, and 5.8 $\mu$m bands and its prominence in the
3.6 and 4.5 $\mu$m bands suggests that the emission is dominated
by the scattered light. The molecular outflow in IRAS 20126 has
been interpreted as driven by a precessing jet based on the
S-shaped locus of the 2.12 $\mu$m H$_2$ knots and the orientation
variation between the inner jet-like SiO outflow and the larger CO
outflow \citep{Cesaroni97, Cesaroni99, Shepherd00, Lebron06,
Su07}. The jet-precessing scenario is also consistent with the
presence of a double system detected at centimeter and
near-infrared wavelengths \citep{Sridharan05, Hofner99, Hofner07}.
Our IRAC observations put new insight into the outflow property of
this luminous source. The IRAC biconical structure far exceeds
inner SiO outflow and covers part of the larger CO outflow. From
the limb-brightening, the cavity wall has PAs from $\sim-33^\circ$
to $\sim-65^\circ$. The PA $\sim-33^\circ$ wall is significantly
offset from the axis of the larger CO outflow which is nearly in a
N-S orientation \citep{Shepherd00, Lebron06}, while the
PA$\sim-65^\circ$ wall has an approximately same orientation as
the jet-like SiO outflow. A natural interpretation of the cavity
(with the limb-brightening) is that it traces a less collimated
biconical component of the IRAS 20126 outflow, rather than being
swept up by a jet precessing from N-S to the current-position of
PA$\sim33^\circ$. However, in contrast to the HH 80-81 outflow (\S
\ref{msf_hh8081}), the jet component in this region coincides with
the wall, not the axis, of the biconical cavity. While multiple
outflows can be one possibility, another possibility can be
derived by adopting the precessing scenario and the theoretical
model unifying the jet-driven and wind-driven low-mass outflows
\citep{Shang07}: when the underlying axial jet and the co-existing
wind are precessing, the dense jet is more efficient in entraining
a collimated molecular outflow via strong shock activities, while
the cavity evacuating by the more tenuous wind component is less
efficient. Consequently, the jet-like SiO outflow can be detected
very close to the current-position of the precessing jet while the
cavity revealed by the scattered light more or less lags behind.

A dense dusty core, which harbors a proto-B star driving the
outflow, was previously detected in the millimeter continuum
\citep{Shepherd00, Cesaroni05}. In the 4.5 and 5.8 $\mu$m bands
the emission feature coincident with the millimeter source is
discernible. In the 8.0 $\mu$m band (Figure \ref{i20126}b), a
point source at this position is detected and the photometry is
listed in Table \ref{irmm}. At higher resolutions
($\lesssim0.5''$), two emission regions oriented along the outflow
axis and separated by a dark lane were detected at near- and
mid-infrared wavelengths \citep{Sridharan05, DeBuizer07}. The
source in the 8.0 $\mu$m band may encompass the previously
detected double sources.

\subsubsection{IRAS 20293+3952} \label{msf_i20293}
In Figure \ref{i20293}, ahead of a previously detected
high-velocity CO outflow \citep{Beuther04b}, a short jet is
detected in the IRAC imaging with an extent of about $10''$. with
the most prominent emission in the 4.5 $\mu$m band. In orientation
the IRAC jet well coincides with the CO outflow. This outflow was
also detected in thermal emissions of SiO and CH$_3$OH
\citep{Beuther04b, Palau07}, which often trace shock activities in
outflows. In the previous 2.12 $\mu$m H$_2$ observations, the
overall structure was not detected but a faint knot can be found
at the southeastern end of the IRAC jet (knot ``C'' in Fig. 2 of
\citet{Kumar02}). We suggest that the IRAC imaging for the first
time reveals the interval driving agent of the CO/SiO/CH$_3$OH
outflow.

A millimeter continuum source, which may harbor an
intermediate-mass protostar, is suggested to be the driving source
of the CO/SiO/CH$_3$OH outflow \citep{Beuther04b, Beuther04c,
Palau07}. The infrared emission from the millimeter source is only
detected in the MIPS 24 $\mu$m band (Table \ref{irmm}). As the
case of AFGL 5142 (\S \ref{msf_afgl5142}), the 24 $\mu$m detection
is located at the bright PSF wing of a nearby saturated source,
resulting in a potential overestimate of the source flux by about
$0.8$ to $0.9$ Jy.

\subsubsection{W75\,N} \label{msf_w75n}
In Figure \ref{w75}, multiple bow-shaped structures in W75\,N can
be identified. The most remarkable one, marked by the double
arrows, form a large scale bipolar bow-shaped structure. The shell
of the southwestern lobe and the tip of the northeastern lobe of
this structure is clearly revealed in the 4.5 $\mu$m band. This
bow-shaped structure has previously been detected in the 2.12
$\mu$m H$_2$ imaging by several authors \citep{Davis98,
Shepherd03, Davis07}, but not reported in previous IRAC
observations toward the DR 21/W75\,N region carried out during the
science verification period of \emph{Spizer} \citep{Persi06,
Davis07}, probably because their observations are far less deep in
integration. The southwestern lobe coincides with the boundary of
the redshifted lobe of a large outflow detected in CO (1-0)
\citep{Shepherd03}, suggesting the bow-shock-driven nature of this
molecular outflow. In addition to this remarkable structure, at
least five bow-shaped structures pointing to the southeast of the
region can be found in Figure \ref{w75}. These structures may be
shocked H$_2$ ``fingers'' tracing multiple bow-shocks driven from
the center of the region, roughly in a fashion similar to the
famous H$_2$ ``fingers'' in the Orion KL region \citep{Schultz99,
Nissen07}. Zooming in on the inner part of this region, the IRAC
imaging reveals a new bow-shaped structure (outlined by a solid
curve in Figure \ref{w75n}). This structure is clearly detected in
the 3.6 and 4.5 $\mu$m bands and appears more prominent in the 4.5
$\mu$m band (Figures \ref{w75n}a, \ref{w75n}b). Its orientation
approximately coincides with the axis of a CO outflow proposed by
\citet{Shepherd03}.

At the center of the large northeast-southwest bow-shaped
structure, \citet{Shepherd01} detected a group of four millimeter
continuum sources (labelled as MM1 to MM4 in Figure \ref{w75n}a).
Embedded in MM1 is a cluster of four centimeter continuum sources,
with three within an area of $1.\!''5$ and another about $3''$ to
the south \citep[Figure \ref{w75n}c,][]{Hunter94, Torrelles97,
Shepherd03}. Of the four centimeter sources, the northernmost one
was interpreted as an ionized radio jet while the other three were
suggested to be UC H{\scriptsize II} regions excited by proto-B
stars. It is still unclear which is driving the large scale CO
outflow. \citet{Shepherd03} suggested the first UC H{\scriptsize
II} region from the north (namely VLA 2) is likely to be the
driving source. The resolution of our IRAC observation does not
allow us to shed any further light on this issue.
\citet{Shepherd03} suggested the second UC H{\scriptsize II}
region from the north (namely VLA 3) to be the driving source of
the CO outflow whose axis approximately coincides with the newly
discovered small bow-shaped structure in the IRAC image. However,
the CO outflow is very compact and confined to within $\sim5''$
from MM1 while the bow-shaped structure is about $50''$ (0.5 pc in
projection) away from the group of the millimeter sources. In
projection MM3 is closer to the axis of this bow-shaped structure
and thus more likely to be the driving source of this structure.
The infrared emission from MM1 is marginally discernible in the
3.6 $\mu$m band (Figure \ref{w75n}a). From 4.5 (Figure
\ref{w75n}b) to 8.0 $\mu$m it is clearly identified as a point
source, and the photometry is listed in Table \ref{irmm}. The MIPS
24 $\mu$m emission from the source is significantly saturated.

\subsubsection{IRAS 22172+5912} \label{msf_i22172}
IRAS 22172+5912 (hereafter IRAS 22172) is the only region in our
sample toward which no outflow signature is found in the IRAC
bands, though a bipolar outflow in CO and HCO$^+$ has been
detected with the OVRO array \citep{Molinari02, Fontani04}. The
peak of the dust emission, which was detected in the OVRO 2.6 mm
and 3.4 mm continuum observations at $\sim4-6''$ resolutions, is
about $5''$ offset from the geometric center of the bipolar
molecular outflow \citep{Molinari02, Fontani04}. The IRAC and MIPS
24$\mu$m imaging do not detect infrared emission from the
millimeter source but reveals two protostars at the center of this
field (see Figure \ref{fig2}i). The southern protostar coincides
with the geometric center of the CO outflow and thus can be a
candidate of the driving source of the outflow. For the millimeter
source, the lack of infrared emission shortward of 24 $\mu$m and
the apparent non-association with the molecular outflow suggest
the source to be at a very young evolutionary stage. Further
(sub)millimeter continuum and line observations with high
sensitivity and higher resolutions are needed to verify whether it
is at a prestellar stage or harbors deeply embedded protostars.

\subsection{UV Heated Reflection Nebulae and Bright Rimmed Clouds}
\label{nebulae} The IRAC imaging of the sample reveals mainly
three classes of extended emission: H$_2$ emission dominated
nebulae most prominent in the 4.5 $\mu$m band (green); scattered
light dominated nebulae most prominent in the 3.6 $\mu$m band
(blue); UV heated hydrocarbon emission dominated reflection
nebulae most prominent in the 8.0 $\mu$m band (red). The first two
classes have been discussed in detail in \S \ref{msf}.

The UV heated reflection nebulae can be found in most regions.
While the first two classes of extended emission are found in
outflows or outflow cavities and thus imply a very young
evolutionary stage of the central source, most of the UV heated
reflection nebulae are associated with young but relatively more
evolved B stars (one exception being the biconical cavity of the
HH 80-81 outflow). For the cometary nebula in AFGL 5142, the
southwestern nebula in IRAS 05358, the southeastern and
northerwestern nebulae in HH 80-81, the central nebula in IRAS
20293, and the northern and central nebulae in W75\,N, a bright
source at the center is detected (denoted as a blue cross in
Figure \ref{fig2}). These sources show infrared excess in our
censes of YSOs based on color-color diagrams, and most of them
were previously suggested to be intermediate- to high-mass young
stars (see annotation of Tabel \ref{photometry}). They are most
likely the exciting sources of the surrounding reflection nebulae.
Of these sources, the one in IRAS 20293 shows an UC H{\scriptsize
II} region and the central one in W75\,N is surrounded by a
relatively extended H{\scriptsize II} region. The spectral types
of the exciting stars of these two H{\scriptsize II} regions,
derived from the Lyman continuum photos required to produce the
observed ionized emission, are B1 and B0.5 for that in IRAS 20293
and W75\,N, respectively \citep{Palau07, Shepherd04b}. Toward the
two sources in HH 80-81, compact centimeter continuum was detected
as well \citep{Marti93}, and the estimated rates of Lyman
continuum photos suggest B3 to B2 stars being the exciting sources
of these two reflection nebulae. There is no reported detection of
a distinct H{\scriptsize II} region surrounding the other
candidate exciting sources. We suggest they are young B stars
capable of providing strong FUV radiation to excite hydrocarbon
emission but too weak EUV to ionize its environment. Spectroscopic
studies are needed to further investigate whether these stars have
circumstellar disks and to confirm their spectral types. There are
also UV heated reflection nebulae for which an exciting source
cannot be identified with the existing data, but the need of FUV
photons to heat hydrocarbons and the lack of significant free-free
emission also suggest a B type star being the exciting source. To
lower the optical depth in FUV and consequently allow the stars to
heat the surrounding hydrocarbons, these B stars must have cleared
most of the surrounding gas. On the other hand, as discussed in \S
\ref{msf}, the central massive star formation site in each region
shows millimeter continuum cores harboring deeply imbedded proto-B
stars. Thus it appears that most regions in the sample are forming
B stars at a range of evolutionary stages.

In two regions, IRAS 20126 and IRAS 22172, the central massive
star formation sites are embedded in a bright large scale
structure which is prominent in the 8.0 $\mu$m band. These two
structures are externally heated rimmed clouds.

\section{Discussion}

\subsection{Clustering} \label{dis_cluster}
One of the distinctive features of massive star formation is that
it occurs in clusters. Consequently, understanding the role of
clustering in the formation of high-mass stars is an essential
step toward a theory of massive star formation. A controversial
issue is whether clusters are necessary for the formation of
high-mass stars, or whether clusters are merely the byproduct of
the formation of high-mass stars in dense massive molecular cores.
In a near-infrared survey of intermediate-mass Herbig Ae/Be stars,
\citet{Testi99} found a smooth transition between the low density
aggregates of young stars associated with stars of spectral type A
or later, and the dense clusters associated with O and early-B
type stars. They argued that the presence of dense clusters is
required for the formation of high-mass stars. The smooth
transition suggested that there may be a fundamental relationship
between the mass of the most massive star and the number and/or
density of stars in a cluster. However, \citet{deWit05} found
evidence that 4\% of O stars form in isolation. This is evidence
that clusters are not necessary for the formation of O stars,
although \citet{Parker07} argued that the isolated O stars may
actually form in small clusters.

To test whether massive stars can form without low-mass stars
requires identification of young massive objects without
associated clusters of low-mass stars; only in these cases, can we
rule out the possibility that an attendant cluster dispersed due
to dynamical evolution. In a $K$-band survey of eight high-mass
star forming cores, \citet{Megeath04} found one core, NGC 6334
I(North), showing evidence of high-mass star formation without
evidence for a cluster of embedded low-mass stars \citep[also see
][]{Megeath99, Hunter06}, although a cluster may still exist
deeply embedded in the cloud \citep{Persi05}. It is important to
search for other potential examples of massive star formation
without the presence of a dense cluster. With its ability to
identify YSOs through infrared excesses and detect deeply embedded
protostars, \emph{Spitzer} is well suited to this task. Although
such regions may yet form clusters, the identification of
high-mass young or protostars without a dense cluster would
demonstrate that low-mass stars are not required for the formation
of massive stars.

Our survey of young intermediate- to high-mass (proto)stars
($\sim10^3$ to $10^5L_{\odot}$) can directly address the role of
clustering in the early stages of intermediate- and high-mass
forming stars. The {\it Spitzer} data provides the ability to
directly identify likely YSOs by the detection of mid-infrared
excesses. The advantage of this approach over previous
near-infrared source counting methods is that the sample does not
significantly suffer from contamination by background stars which
can be significant for high-mass star forming regions in the
Galactic plane \citep{Pratap99}. Also, the mid-infrared colors are
much more sensitive to protostellar objects, and candidate
protostellar objects can be identified through their colors.
However, the bright mid-infrared nebulosity in high-mass star
forming regions, the $\sim2-3''$ angular resolution of the IRAC
data, and the modest sensitivity of the 2MASS photometry limit the
detection of sources, in particular in regions of bright
nebulosity. We expect the census of YSOs to be incomplete in all
of the regions. Furthermore, we expect that a certain fraction of
the stars do not have disks and will not be identified in our
analysis; the fraction of stars without disks is approximately
30\% for 1 Myr old low-mass stars \citep{Hernandez07}.

We group the objects by their total far-infrared luminosity, as
measured from the IRAS point source catalog; this luminosity will
be dominated by the most massive objects in the regions. In all
the regions, associated low-mass YSOs are detected.  The
$\sim10^3L_{\odot}$ regions, AFGL 5142, G192, IRAS 05358 and IRAS
20293, show parsec sized clusters with around 20 YSOs surrounded
by a more extended and sparse distribution of young stars and
protostars. The $\sim10^4L_{\odot}$ regions IRAS 19410 and HH
80-81 also show clusters and extended components. Except AFGL
5142, the clusters associated with these regions do not show
concentrations of stars toward the central massive objects. This
may be due to the decrease in completeness toward the bright
nebulosities in the central regions. Finally, the
$10^{5.1}L_{\odot}$ region W75\,N shows a cluster that is both
richer and more spatially extended. This region also shows a
paucity of sources in the bright nebulous center of the region,
further indicating the observations are significantly limited by
completeness.

In Figure \ref{fig17} we plot the number of sources as a function
of the source luminosity. To mitigate the effects of
incompleteness on our samples, we plot the number of sources with
3.6 $\mu$m magnitudes lower than 13 and 14. Given a typical
$Ks-[3.6]$ color of 0.25, an extinction of $A_K\sim0.25$ at 1.8
kpc, and an age of 1 Myr, these limits correspond to masses of 1.2
and 0.5 $M_{\odot}$ \citep{Baraffe98}. The two plots show a trend
of increasing number of associated YSOs with increasing total
luminosity. In particular, the number of YSOs in the
$10^{5.1}L_{\odot}$ region (W75\,N) is significant higher than
those in $\sim10^4L_{\odot}$ regions, which again are somewhat
higher in number than the $\sim10^3L_{\odot}$ regions. It is
possible that a trend of increasing incompleteness with higher
far-infrared luminosities is decreasing the slope of this trend.

There are two main exceptions to this trend: IRAS 22172 and IRAS
20126. IRAS 22172 appears to be part of a larger, potentially more
evolved region which has undergone significant gas clearing. The
clearing of the gas would also lower the fraction of light
absorbed by dust and re-emitted in the infrared; hence this region
shows a large number of stars for its measured IRAS far-infrared
luminosity. The other exception is IRAS 20126, which has the
lowest number of associated YSOs in the sample. Unlike the other
regions, IRAS 20126 shows no obvious cluster in the field. This is
unlikely to be solely the effects of incompleteness, with the
average signal of about 2.5 MJy/sr at 4.5 $\mu$m and 52 MJy/sr at
8.0 $\mu$m in the central parsec area, compared to 6.5 MJy/sr at
4.5 $\mu$m and 67 MJy/sr at 8.0 $\mu$m for HH 80-81 and 3.8 MJy/sr
at 4.5 $\mu$m and 72 MJy/sr at 8.0 $\mu$m for IRAS 19410.

More rigorous studies of this sample await deeper near-infrared
data to complement the {\it Spitzer} 3.6 and 4.5 $\mu$m imaging.
However, the analysis here already illustrates several features.
First, there does seem to be a trend of the number of associated
YSOs with luminosity. Second, these regions show both clusters in
the central 1 pc as well as more extended halos of sources around
the clusters. Finally, there does seem to be a significant
dispersion in the number of associated YSOs for a given total
far-infrared luminosity, with one luminous source, IRAS 20126,
showing no obvious cluster. The lack of a cluster warrants future
work; if confirmed this would demonstrate that the formation of
$\sim10^4L_{\odot}$ sources is not dependent on the presence of
deep clusters.

\subsection{Outflows and Outflow Cavities}
For a sample of nine high-mass star forming regions associated
with CO and/or SiO outflows, the deep IRAC imaging reveals
outflows and outflow cavities toward eight regions, illustrating
that the IRAC imaging can be an effective tool for outflow
detection. For the jets in IRAS 05358, AFGL 5142, IRAS 20293, the
bow-shock shells in W75\,N, and the nebulosities in the eastern
lobe of the G192 outflow, the prominence in the 4.5 $\mu$m band
and coincidence with the ground-based 2.12 $\mu$m H$_2$ detections
(if detected in the 2.12 $\mu$m H$_2$ line) suggest the emission
is mostly attributed to shocked H$_2$ emission, although the
contribution from H{\scriptsize I} Br$\alpha$ (4.052 $\mu$m), CO
v=1-0 (4.45-4.95 $\mu$m) lines cannot be ruled out without
spectroscopic observations. A detailed investigation of the
emission in the biconical cavity in IRAS 20126 shows strong
evidence for the scattered light being dominant in the 3.6 and 4.5
$\mu$m bands (Qiu et al., in prep.). The elliptical structure in
IRAS 19410 appears relatively diffuse and most prominent in the
3.6 $\mu$m band. The emission in this outflow can be dominated by
the scattered light as well. The clear prominence of the biconical
structure in HH 80-81 suggested is dominated by UV heated
hydrocarbon emission.

\subsubsection{An evolutionary picture of massive outflows?}
Both theoretical models and high spatial resolution observations
suggest an evolutionary scenario for low-mass outflows: as
low-mass stars evolve from class 0 through class I to class II
stages, the mass loading in the winds and density structure in the
cores conspire the widening of outflows \citep{Fuller02, Arce06,
Shang07}. While for massive outflows, whether an evolutionary
picture exists is unknown \citep{Beuther05}.

The detected outflows in the sample show dramatically different
morphologies. We detect highly collimated jets, bow-shock shells,
and biconical cavities. Given the extreme complexity of the CO
observations \citep{Beuther03} and relatively confused emission in
the IRAC imaging compared with the other regions, we leave out the
IRAS 19410 outflow in the following discussion. Considering Jet1
in IRAS 05358, the short jet in AFGL 5142, and the jet in IRAS
20293, the IRAC observations reveal the internal driving agents of
well collimated CO outflows in these regions. The powering sources
of Jet1 in IRAS 05358 and the short jet in AFGL 5142 are suggested
to be deeply embedded proto-B stars, toward which only very weak
radio continuum was detected \citep{Beuther07,Zhang07}. The jet in
IRAS 20293 is most likely driven by an intermediate-mass protostar
showing no detectable centimeter emission \citep{Beuther04b}. In
contrast, the large scale CO outflows in W75\,N and G192 appear to
be poorly collimated within $\lesssim0.5$ pc from the central
driving sources. In W75\,N, proper-motion observations of water
masers by \citet{Torrelles03} delineated a non-collimated, shell
outflow at a 160 AU scale expanding in multiple directions with
respective to the UC H{\scriptsize II} region VLA 2, which may
drive the large scale CO outflow as suggested by
\citet{Shepherd03}. The bipolar bow-shaped structure beyond
$\sim1$ pc detected in the 2.12 $\mu$m H$_2$ and the IRAC 4.5
$\mu$m imaging could be the remanent of a collimated component.
For the G192 outflow, no collimated component can be found within
$\sim2.4$ pc from the driving source. The proposed central driving
sources of these two poorly collimated outflows are found to be
surrounded by UC H{\scriptsize II} regions. If we adopt the
centimeter free-free emission as an indicator of the relative
evolution between these sources, there seems to be a trend of
evolution for the related outflows: both internal driving agents
and entrained molecular outflows appear to be highly collimated
for the youngest sources (e.g., IRAS 05358, AFGL 5142, IRAS
20293); as the central source evolves to form a significant UC
H{\scriptsize II} region, only poorly collimated outflow
structures can be found around the central driving source (e.g.,
W75\,N, G192).

For the IRAS 20126 outflow, it remains ambiguous whether it is
merely composed of a precessing jet and consequently larger
structures are all entrained/swept-up by this jet or it has a
jet-like component as well as a biconical component with a
moderate opening angle. Toward the central driving source of this
outflow, very weak centimeter continuum was detected, presumably
suggesting the source being at a very young evolutionary stage.
For the HH 80-81 outflow, our \emph{Spitzer} data and previous
centimeter continuum reveal a remarkable two-component outflow
with a biconical cavity surrounding an axial radio jet. Strong
centimeter continuum was detected toward the central driving
source of the HH 80-81 outflow, presumably suggesting the source
being relatively more evolved compared with that of IRAS 20126.
The outflow cavity of HH 80-81 also shows emission in the
mid-infrared hydrocarbon features, suggesting that UV radiation
from the central sources are heating the cavity walls. Therefore,
if the IRAC biconical cavity in IRAS 20126 is a complementary less
collimated component of the previously detected SiO jet, there
will be a large evolutionary time scale for the existence of both
a highly collimated jet-like component and a less collimated
biconical component for outflows from proto-B stars. With the
existing data it is very difficult to determine whether the
central source of IRAS 20126 is more evolved than those of AFGL
5142 and IRAS 05358, and whether the central source of HH 80-81 is
younger than those of W75\,N and G192. Presumably it is likely
that the IRAS 20126 and HH 80-81 outflows, with the former being
younger, represent evolutionary stages between that of the AFGL
5142 \& IRAS 05358 outflows, for which both internal driving
agents and entrained gas appear well collimated, and that of the
W75\,N \& G192 outflows, for which only poorly collimated
structures can be found around the central driving sources.

However, such a trend is only a tentative interpretation based on
current observations of a small sample. The question whether there
is an evolutionary picture for massive outflows is far from
conclusive, and remains an observational challenge. In high-mass
star forming regions, multiple outflows with complicated
structures are often detected. Due to relatively larger distances
and crowded clustering mode, it is very difficult with the current
facilities to reliably identify the driving source of a massive
outflow. Also resolving the launching zone of a massive outflow,
which is of great importance for investigate driving mechanisms,
requires extremely high spatial resolution and high sensitivity.
In addition, observations of outflows in more luminous objects
($L\,{\gtrsim}\,10^5L_{\odot}$), in particular high spatial
resolution observations, are still rare (the combined bolometric
luminosity of UC H{\scriptsize II} regions in MM-1 in W75\,N is
$10^{4.6}L_{\odot}$). Extensive and comprehensive studies of
outflows in this luminosity regime would be crucial for testing
possible evolutionary scenarios of massive outflows and
understanding formation processes of O type stars.

\section{Summary}
We have described initial results of mid-infrared imaging
observations toward nine high-mass star forming regions made with
the IRAC and MIPS cameras onboard \emph{Spitzer}. The regions were
selected for the presence of luminous ($> 10^3$~L${_\odot}$) young
objects driving molecular outflows. The observed fields are
approximately $5'\times5'$, corresponding to physical widths of
$\sim2.5-3.5$ pc for the source distances ranging from
$\sim1.7-2.4$ kpc.

Using the $3-24\,\mu$m \emph{Spitzer} photometry in combination
with near-infrared 2MASS data, we identify a total of 417 YSOs
with infrared excesses attributed to circumstellar disks or
envelopes, including at least 12 candidates of intermediate- to
high-mass young stars. In most regions, the spatial distribution
of these YSOs shows both a cluster component centered on the sites
of massive star formation as well as a more extended distribution
of YSOs outside the cluster. The number of YSOs grows with the
total far-infrared luminosity, with two significant exceptions:
IRAS 22172 appears to be part of a larger, more evolved region;
IRAS 20126 appears to be a $10^4$\,L$_{\odot}$ protostar with only
19 associated YSOs and no central cluster.

Zooming in on the central massive star formation sites, we search
for infrared counterparts toward millimeter continuum sources
identified in interferometer observations. We detected
counterparts for eight millimeter continuum sources in the IRAC
bands or MIPS 24 $\mu$m band. These eight sources appear to be
natal cores of proto-B stars driving molecular outflows. All of
the regions of massive star formation have bright associated
nebulosity from scattered light, shocked H$_2$ and/or UV heated
hydrocarbon emission features. This nebulosity limits the
completeness of our YSO survey in the central areas.

The deep IRAC imaging detects features associated with 12 outflows
in eight of the surveyed regions. Compared with previous
ground-based observations, our IRAC observations confirm previous
2.12 $\mu$m H$_2$ jets or bow-shock shaped structures in AFGL
5142, IRAS 05358, and W75\,N (the large bow-shock shaped
structure), and previous H${\scriptsize \alpha}$/[S{\scriptsize
II}] features in G192; reveal new structures in HH 80-81, IRAS
20126, and IRAS 19410; detect a new H$_2$ jet in IRAS 20293 and a
new bow-shock shaped structure in W75\,N (the small bow-shock
shaped structure). In one case, HH 80-81, the inner surface of the
outflow cavity shows strong 8 $\mu$m emission indicative of
hydrocarbons heated by UV radiation. Based on the morphological
variations of the detected outflows in conjunction with previous
observations, we outline a possible evolutionary picture for
massive outflows.

UV heated reflection nebulae dominated by hydrocarbon emission in
the 8.0 $\mu$m band can be found in most regions. They may imply
the presence of relatively more evolved young B stars. Externally
heated bright rimmed clouds are found in IRAS 20126 and IRAS
22172.

\acknowledgments We are grateful to Luis F. Rodr\'{i}guez for
providing us the VLA cm data on HH 80-81. This work is based on
observations made with the \emph{Spitzer} Space Telescope, which
is operated by the Jet Propulsion Laboratory, California Institute
of Technology under a contract with the National Aeronautics and
Space Administration (NASA). This publication makes use of data
products form the Two Micron All Sky Survey, which is a joint
project of the University of Massachusetts and the Infrared
Processing and Analysis Center/California Institute of Technology,
funded by NASA and the National Science Foundation. K. Q.
acknowledges the support of the Grant 10128306 from NSFC. H. B.
acknowledges financial support by the Emmy-Noether-Program of the
Deutsche Forschungsgemeinschaft (DFG, grant BE2578).

\clearpage

\begin{deluxetable}{cccccc}
\tablewidth{0pc} \tabletypesize{\scriptsize} \tablecaption{Summary
of the Sample \label{source}}

\tablehead{ \colhead{} & \multicolumn{2}{c}{Reference Center} &
\colhead{IRAS Luminosity} & \colhead{Distance} & \colhead{} \\
\cline{2-3} \\[-2ex] \raisebox{2.3ex}[0pt]{Source Name} &
\colhead{RA(J2000)} & \colhead{DEC(J2000)} &
\colhead{($L_{\odot}$)} & \colhead{(kpc)} &
\raisebox{2.3ex}[0pt]{Number of Identified YSOs}}

\startdata
AFGL 5142         &  05 30 48.00  & +33 47 53.8  & $10^{3.4}$ &  1.8 & 44 \\
IRAS 05358+3543   &  05 39 12.78  & +35 45 50.6  & $10^{3.8}$ &  1.8 & 30 \\
G192.16-3.82      &  05 58 13.52  & +16 31 58.3  & $10^{3.4}$ &  2.0 & 31 \\
HH 80-81          &  18 19 12.09  & -20 47 30.9  & $10^{4.2}$ &  1.7 & 48 \\
IRAS 19410+2336   &  19 43 11.00  & +23 44 35.0  & $10^{4.0}$ &  2.1 & 46 \\
IRAS 20126+4104   &  20 14 26.00  & +41 13 32.0  & $10^{4.0}$ &  1.7 & 19 \\
IRAS 20293+3952   &  20 31 10.70  & +40 03 10.0  & $10^{3.8}$ &  2.0 & 31 \\
W75\,N            &  20 38 36.45  & +42 37 34.0  & $10^{5.1}$ &  2.0 & 77 \\
IRAS 22172+5549   &  22 19 08.60  & +56 05 02.0  & $10^{3.3}$ &  2.4 & 91 \\
\enddata
\end{deluxetable}

\clearpage

\begin{deluxetable}{ccccccccccc}
\tablewidth{0pc} \tabletypesize{\scriptsize} \rotate
\tablecaption{Point Source Detections and Limiting Magnitudes
\label{observation}}

\tablehead{ \colhead{} & \multicolumn{2}{c}{3.6 $\mu$m} &
\multicolumn{2}{c}{4.5 $\mu$m} & \multicolumn{2}{c}{5.8 $\mu$m}
& \multicolumn{2}{c}{8.0 $\mu$m} & \multicolumn{2}{c}{24 $\mu$m} \\
\cline{2-3} \cline{4-5} \cline{6-7} \cline{8-9} \cline{10-11} \\[-2ex]
\raisebox{2.3ex}[0pt]{Source Name} & \colhead{number} &
\colhead{limiting mag} & \colhead{number} & \colhead{limiting mag}
& \colhead{number} & \colhead{limiting mag} & \colhead{number} &
\colhead{limiting mag} & \colhead{number} & \colhead{limiting mag}
}

\startdata
AFGL 5142         &  517  & 18.206$\pm$0.166 & 545 & 17.790$\pm$0.168 & 193 & 15.849$\pm$0.170 & 128& 15.073$\pm$0.215 & 16 & 9.774$\pm$0.131 \\
IRAS 05358+3543   &  377  & 17.674$\pm$0.120 & 410 & 17.522$\pm$0.185 & 122 & 15.168$\pm$0.187 & 76 & 13.973$\pm$0.124 & 12 & 8.386$\pm$0.148 \\
G192.16-3.82      &  432  & 18.365$\pm$0.158 & 449 & 17.788$\pm$0.156 & 130 & 15.640$\pm$0.199 & 75 & 15.271$\pm$0.200 & 14 & 9.850$\pm$0.084 \\
HH 80-81          &  783  & 16.529$\pm$0.129 & 862 & 16.057$\pm$0.193 & 578 & 15.066$\pm$0.217 & 402& 13.930$\pm$0.239 & 17 & 7.821$\pm$0.184 \\
IRAS 19410+2336   &  750  & 16.385$\pm$0.127 & 844 & 16.181$\pm$0.119 & 411 & 14.807$\pm$0.181 & 203& 13.849$\pm$0.216 & 21 & 7.396$\pm$0.067 \\
IRAS 20126+4104   &  630  & 17.070$\pm$0.146 & 808 & 16.942$\pm$0.155 & 253 & 15.894$\pm$0.241 & 154& 14.807$\pm$0.196 &  7 & 7.834$\pm$0.174 \\
IRAS 20293+3952   &  646  & 16.977$\pm$0.130 & 769 & 16.609$\pm$0.159 & 282 & 14.953$\pm$0.198 & 137& 13.417$\pm$0.159 & 11 & 7.932$\pm$0.133 \\
W75\,N            &  627  & 17.077$\pm$0.140 & 798 & 16.869$\pm$0.194 & 227 & 14.371$\pm$0.235 & 115& 12.732$\pm$0.208 & 11 & 7.894$\pm$0.144 \\
\enddata
\end{deluxetable}

\clearpage

\begin{deluxetable}{ccccccccccc}
\tablewidth{0pc} \tabletypesize{\scriptsize} \rotate
\tablecaption{YSOs: 2MASS, IRAC, MIPS 24 $\mu$m Photometry
\label{photometry}}

\tablehead{ \colhead{RA\tablenotemark{\dag}} &
\colhead{DEC\tablenotemark{\dag}} & \colhead{$J$} & \colhead{$H$}
& \colhead{$Ks$} & \colhead{[3.6]} & \colhead{[4.5]} &
\colhead{[5.8]} & \colhead{[8.0]} &
\colhead{[24]\tablenotemark{\ddag}} &
\colhead{Remark\tablenotemark{\ast}} }

\startdata
%\cutinhead{AFGL 5142}
            &             &      &      &      &   AFGL 5142      &                  &                  &                  &      &    \\
\tableline
05 30 36.61 & +33 49 18.9 &  --- &  --- &  --- & 14.945$\pm$0.013 & 14.597$\pm$0.015 & 14.330$\pm$0.051 & 13.522$\pm$0.073 &  --- & II \\
05 30 36.79 & +33 47 18.7 &  --- &  --- &  --- & 14.672$\pm$0.014 & 14.345$\pm$0.011 & 13.975$\pm$0.053 & 13.331$\pm$0.046 &  --- & II \\
05 30 37.98 & +33 49 02.2 &  --- &  --- &  --- & 11.907$\pm$0.006 & 11.530$\pm$0.005 & 11.264$\pm$0.006 & 10.540$\pm$0.021 &  7.149$\pm$0.021 & II \\
05 30 40.18 & +33 46 18.2 &  --- &  --- &  --- & 15.257$\pm$0.016 & 14.343$\pm$0.011 & 13.485$\pm$0.029 & 12.623$\pm$0.029 &  9.156$\pm$0.073 & I \\
05 30 42.14 & +33 48 32.8 &  --- &  --- &  --- & 15.162$\pm$0.035 & 14.426$\pm$0.036 & 13.428$\pm$0.060 & 12.624$\pm$0.061 &  --- & II \\
05 30 42.36 & +33 48 15.1 & 12.523$\pm$0.023 & 11.502$\pm$0.022 & 10.905$\pm$0.019 &  9.896$\pm$0.002 &  9.407$\pm$0.002 &  8.877$\pm$0.002 &  8.064$\pm$0.002 &  4.982$\pm$0.035 & II \\
05 30 42.67 & +33 48 59.7 &  --- &  --- &  --- & 14.783$\pm$0.012 & 14.573$\pm$0.015 & 14.510$\pm$0.067 & 14.011$\pm$0.099 &  --- & II \\
\enddata
\end{deluxetable}

\clearpage

\begin{deluxetable}{ccc}
\tablewidth{0pc} \tabletypesize{\scriptsize}
\tablecaption{Completeness Estimates\label{fraction}}

\tablehead{\colhead{Source Name} &
\colhead{[3.6]\tablenotemark{a}} &
\colhead{Completeness\tablenotemark{b}}}

\startdata
AFGL 5142         &  15  & 84\% \\
IRAS 05358+3543   &  14.5  & 79\% \\
G192.16-3.82      &  15  & 81\% \\
HH 80-81          &  13.5  & 84\% \\
IRAS 19410+2336   &  13.5  & 73\% \\
IRAS 20126+4104   &  14.5  & 73\% \\
IRAS 20293+3952   &  13  & 85\% \\
W75\,N            &  12.5  & 75\% \\
IRAS 22172+5549   &  14.5  & 76\% \\
\enddata
\tablenotetext{a}{Completeness magnitudes at 3.6 $\mu$m derived
from Figure \ref{complete}.} \tablenotetext{b}{For sources with
the 3.6 $\mu$m photometry brighter than the completeness
magnitude, the fraction of the sources that can be placed on one
or more color-color diagrams.}
\end{deluxetable}

\clearpage

\begin{deluxetable}{ccccccccc}
\tablewidth{0pc} \tabletypesize{\scriptsize} \rotate
\tablecaption{Infrared Counterparts of Millimeter Cores
\label{irmm}}

\tablehead{\colhead{Source} & \colhead{R.A. (J2000)} &
\colhead{Decl. (J2000)} & \colhead{[3.6]} & \colhead{[4.5]} &
\colhead{[5.8]} & \colhead{[8.0]} & \colhead{[24]} &
\colhead{Ref.\tablenotemark{\dag}} }

\startdata
AFGL 5142         &  05 30 47.94  & +33 47 54.2  & --- &  --- & --- & --- & 0.632$\pm$0.005 & 1 \\
IRAS 05358+3543   &  05 39 13.04  & +35 45 51.2  & --- & 7.582$\pm$0.020 & 6.250$\pm$0.015 & 5.098$\pm$0.011 & $\ll$0.626 & 2, 3 \\
G192.16-3.82      &  05 58 13.55  & +16 31 58.2  & 9.063$\pm$0.009 & 7.726$\pm$0.007 & 6.915$\pm$0.007 & 5.954$\pm$0.014 & $\ll$0.626 & 4, 5 \\
HH 80-81          &  18 19 12.11  & -20 47 31.0  & --- & --- & 4.104$\pm$0.050 & 3.040$\pm$0.081 & $\ll$0.626 & 6 \\
IRAS 19410+2336   &  19 43 11.25  & +23 44 03.2  & 6.682$\pm$0.018 & 4.624$\pm$0.008 & 3.365$\pm$0.004 & 2.780$\pm$0.008 & $\ll$0.626 & 7, 8 \\
IRAS 20126+4104   &  20 14 25.97  & +41 13 33.1  & --- & --- & --- & 5.422$\pm$0.060 & $\ll$0.626 & 9, 10 \\
IRAS 20293+3952   &  20 31 12.92  & +40 03 23.0  & --- & --- & --- & --- & 1.123$\pm$0.021 & 11, 12, 13 \\
W75\,N            &  20 38 36.49  & +42 37 33.5  & --- & 6.161$\pm$0.22 & 4.451$\pm$0.017 & 3.346$\pm$0.018 & $\ll$0.626 & 14 \\
\enddata
\tablenotetext{\dag}{References of interferometric millimeter
continuum observations: 1. \citet{Zhang07}; 2. \citet{Beuther02};
3. \citet{Beuther07}; 4. \citet{Shepherd99}; 5.
\citet{Shepherd04a}; 6. \citet{Gomez03}; 7. \citet{Beuther03}; 8.
\citet{Beuther04a}; 9. \citet{Shepherd00}; 10. \citet{Cesaroni05};
11. \citet{Beuther04b}; 12. \citet{Beuther04c}; 13.
\citet{Palau07}; 14. \citet{Shepherd01}}
\end{deluxetable}

\clearpage

\clearpage

\begin{figure}
\plotone{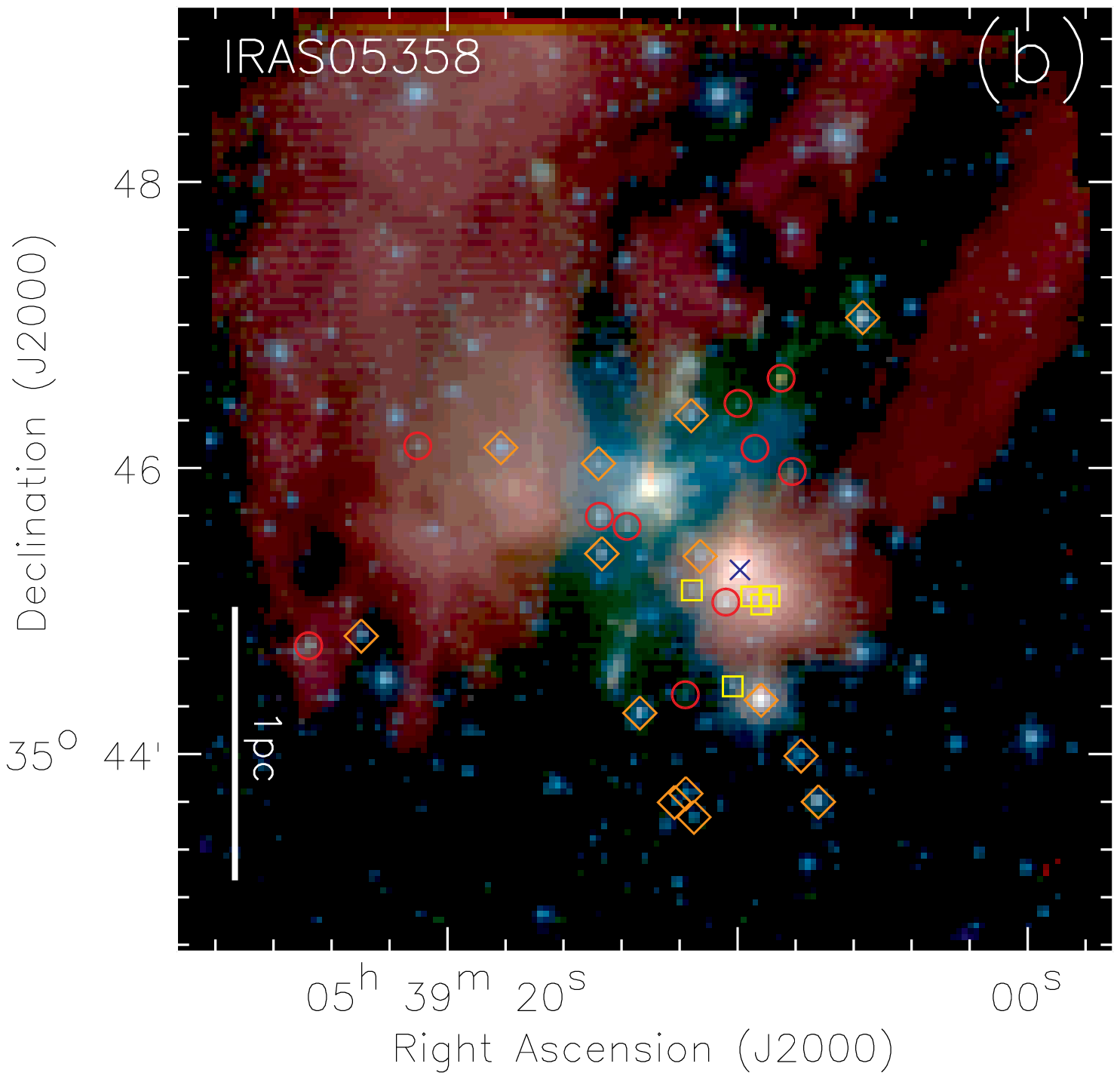}\\
Fig. \ref{fig2}. --- Continued
\end{figure}

\begin{figure}
\plotone{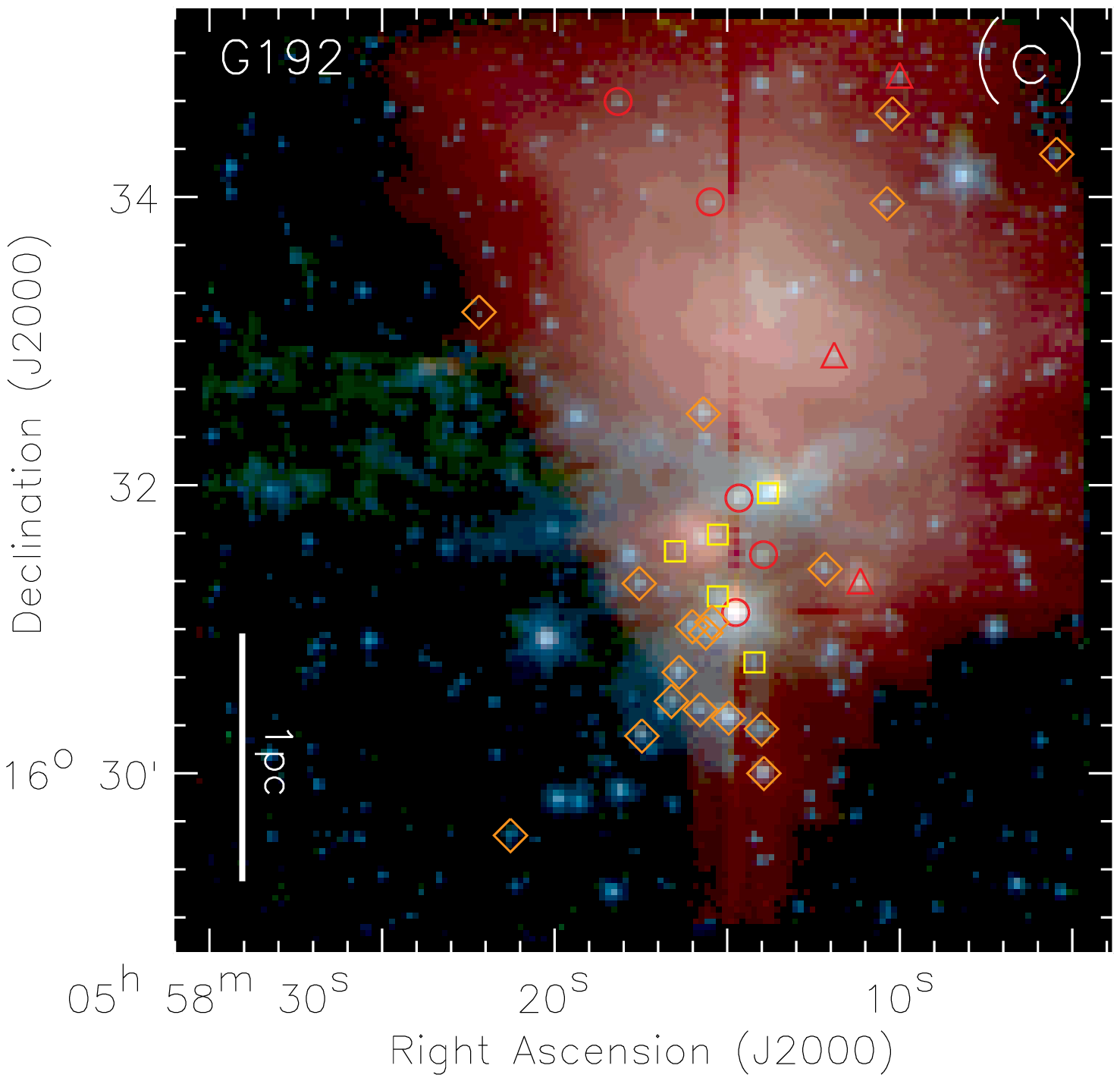}\\
Fig. \ref{fig2}. --- Continued
\end{figure}

\begin{figure}
\plotone{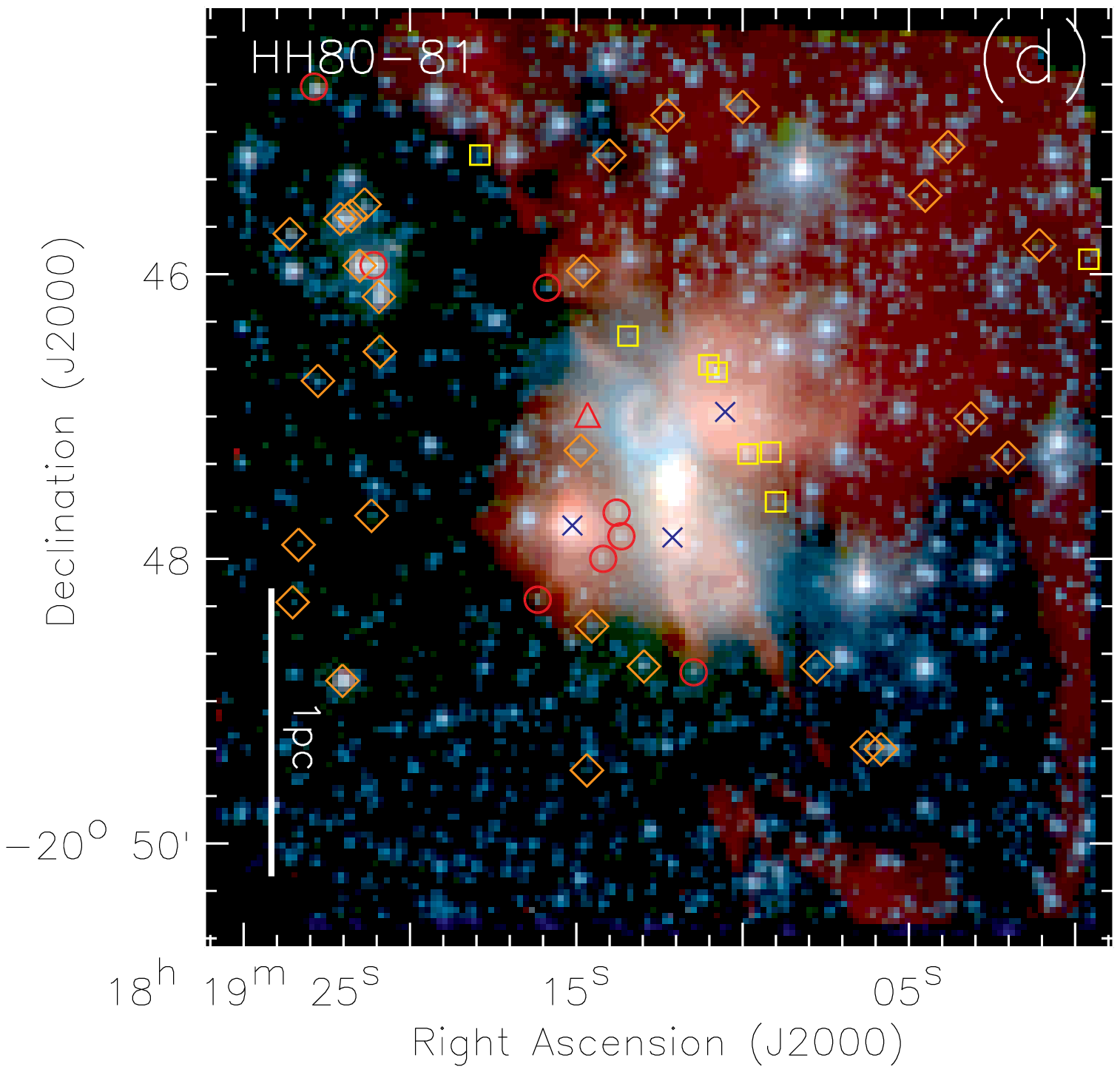}\\
Fig. \ref{fig2}. --- Continued
\end{figure}

\begin{figure}
\plotone{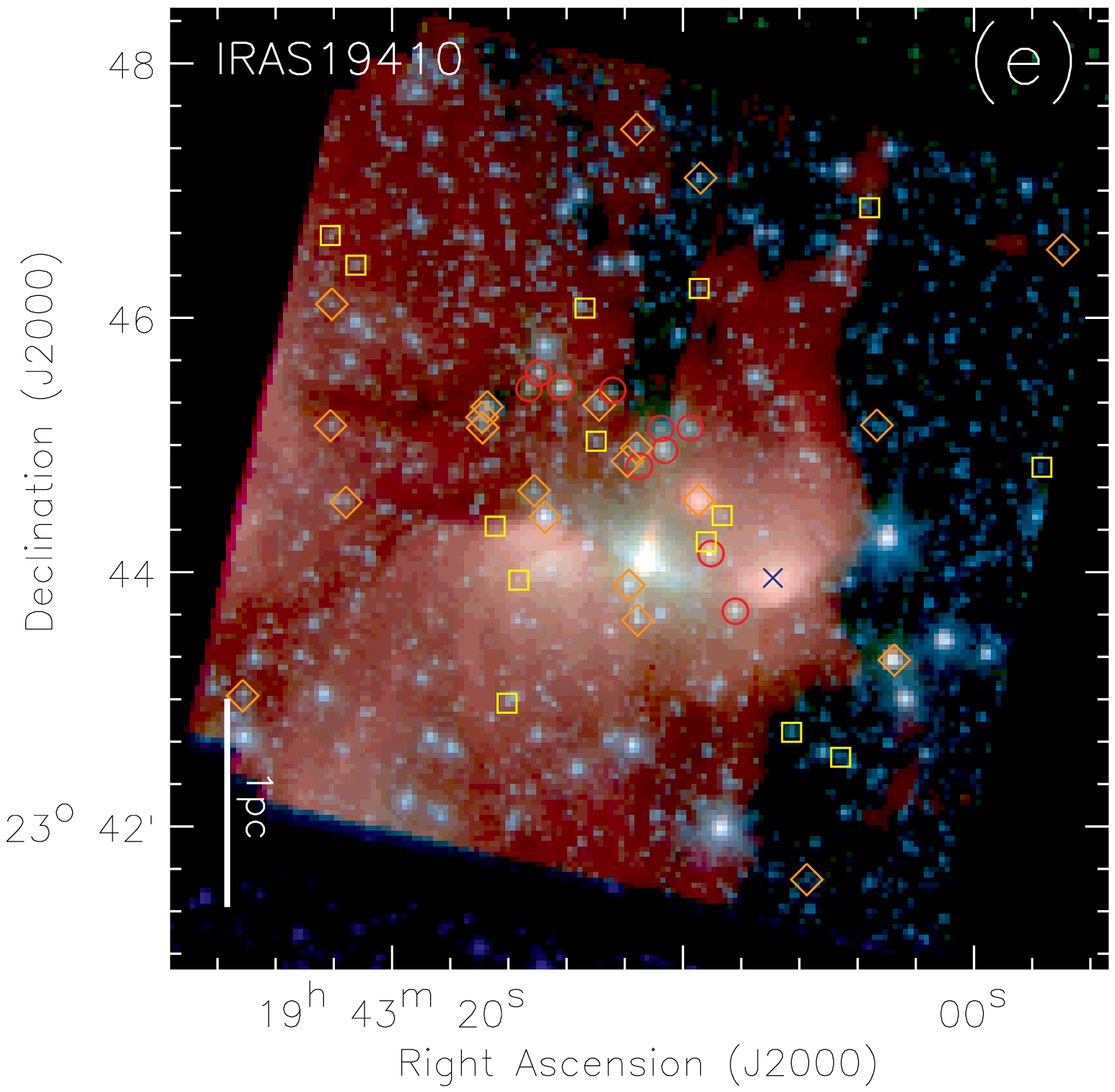}\\
Fig. \ref{fig2}. --- Continued
\end{figure}

\begin{figure}
\plotone{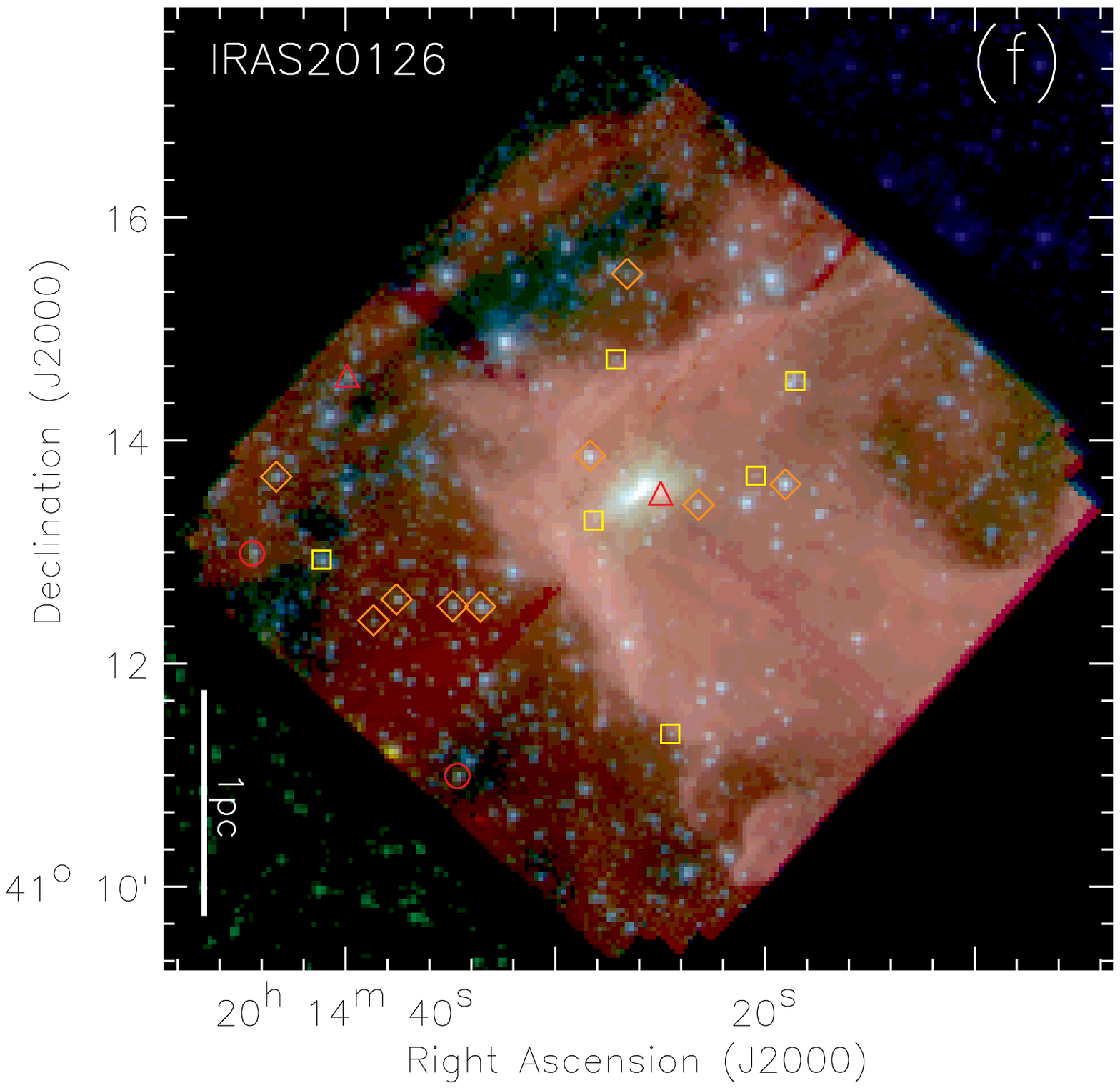}\\
Fig. \ref{fig2}. --- Continued
\end{figure}

\begin{figure}
\plotone{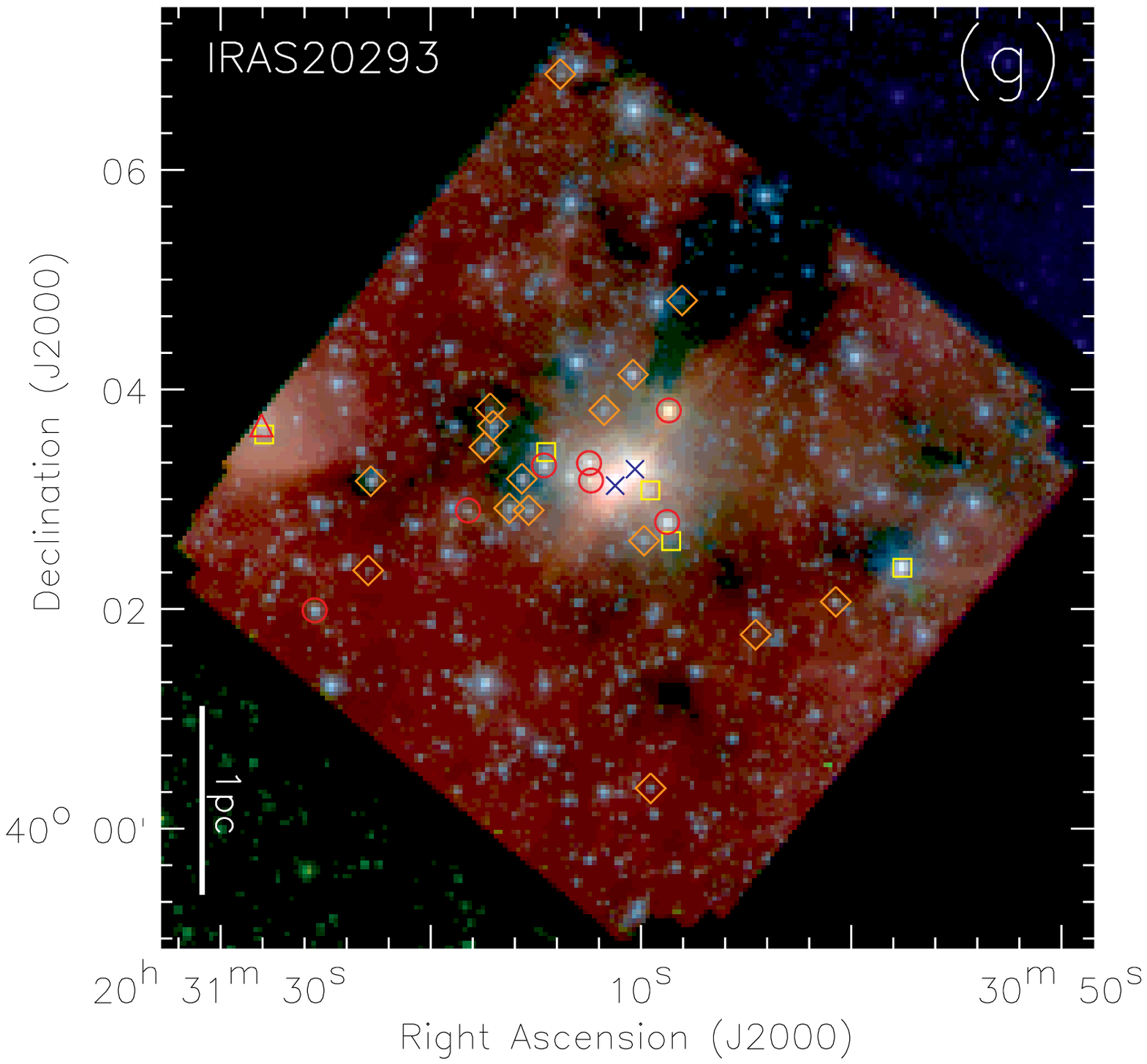}\\
Fig. \ref{fig2}. --- Continued
\end{figure}

\begin{figure}
\plotone{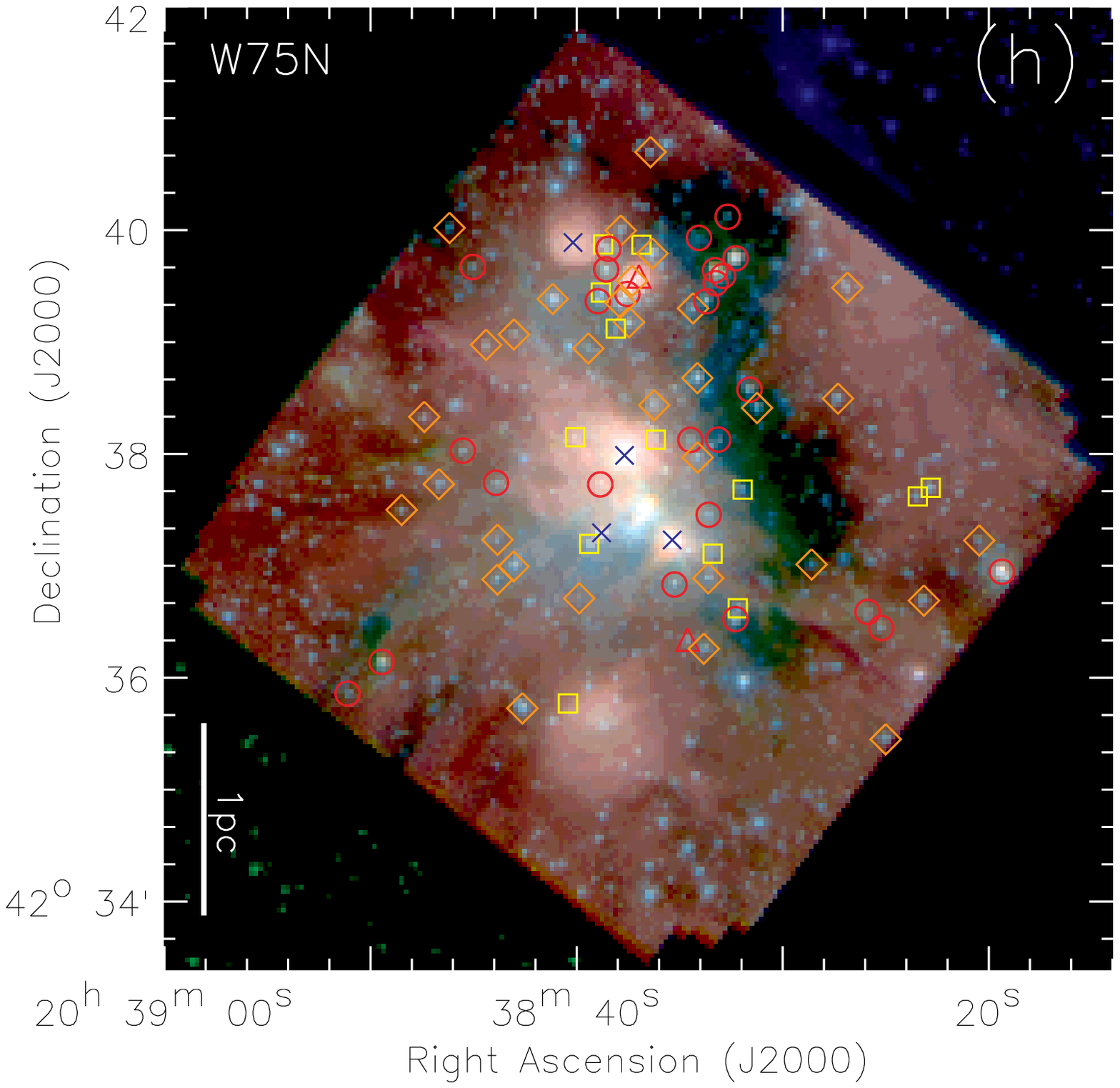}\\
Fig. \ref{fig2}. --- Continued
\end{figure}

\begin{figure}
\plotone{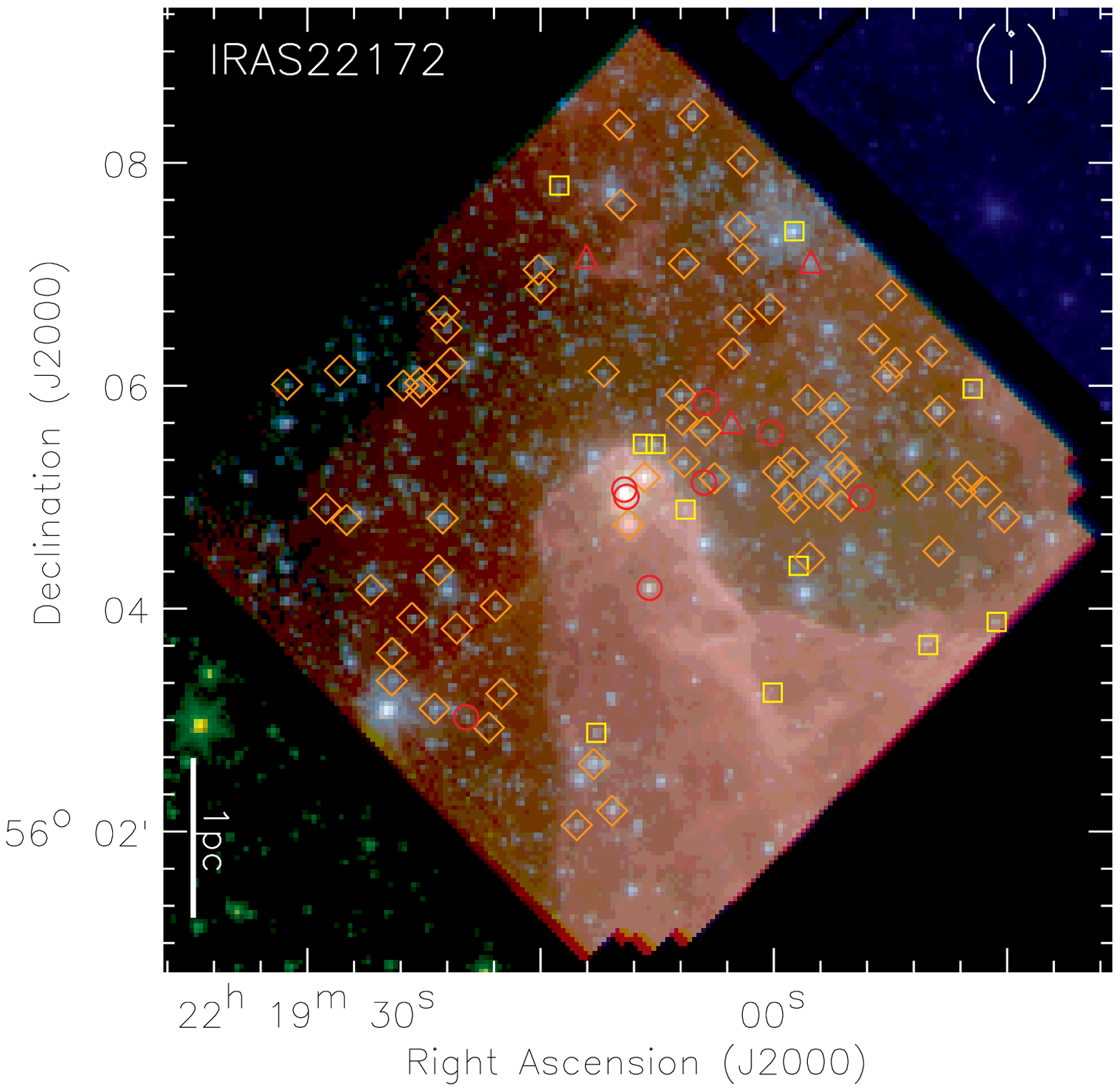}\\
Fig. \ref{fig2}. --- Continued
\end{figure}

\clearpage

\begin{figure}
\epsscale{.32} \plotone{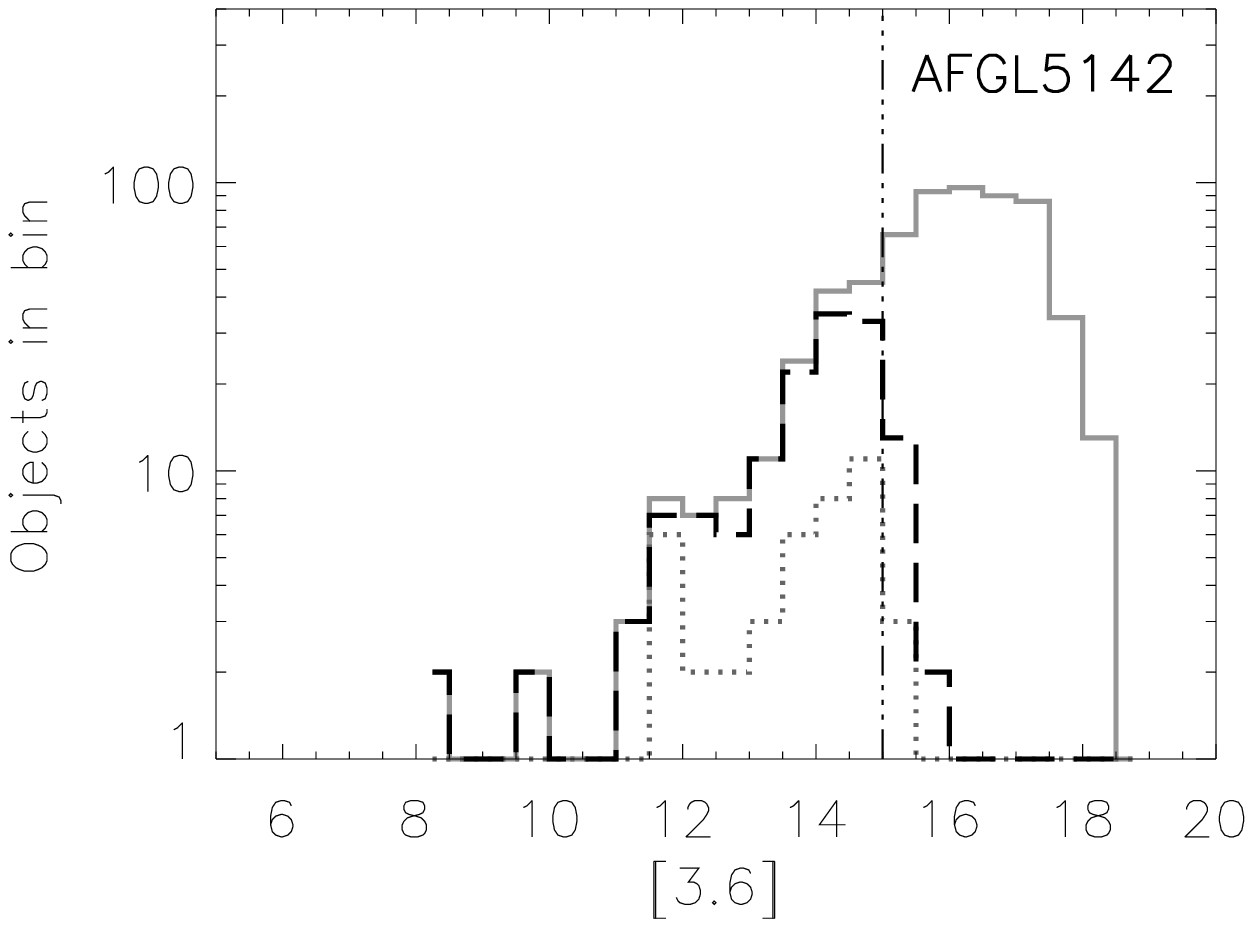}\plotone{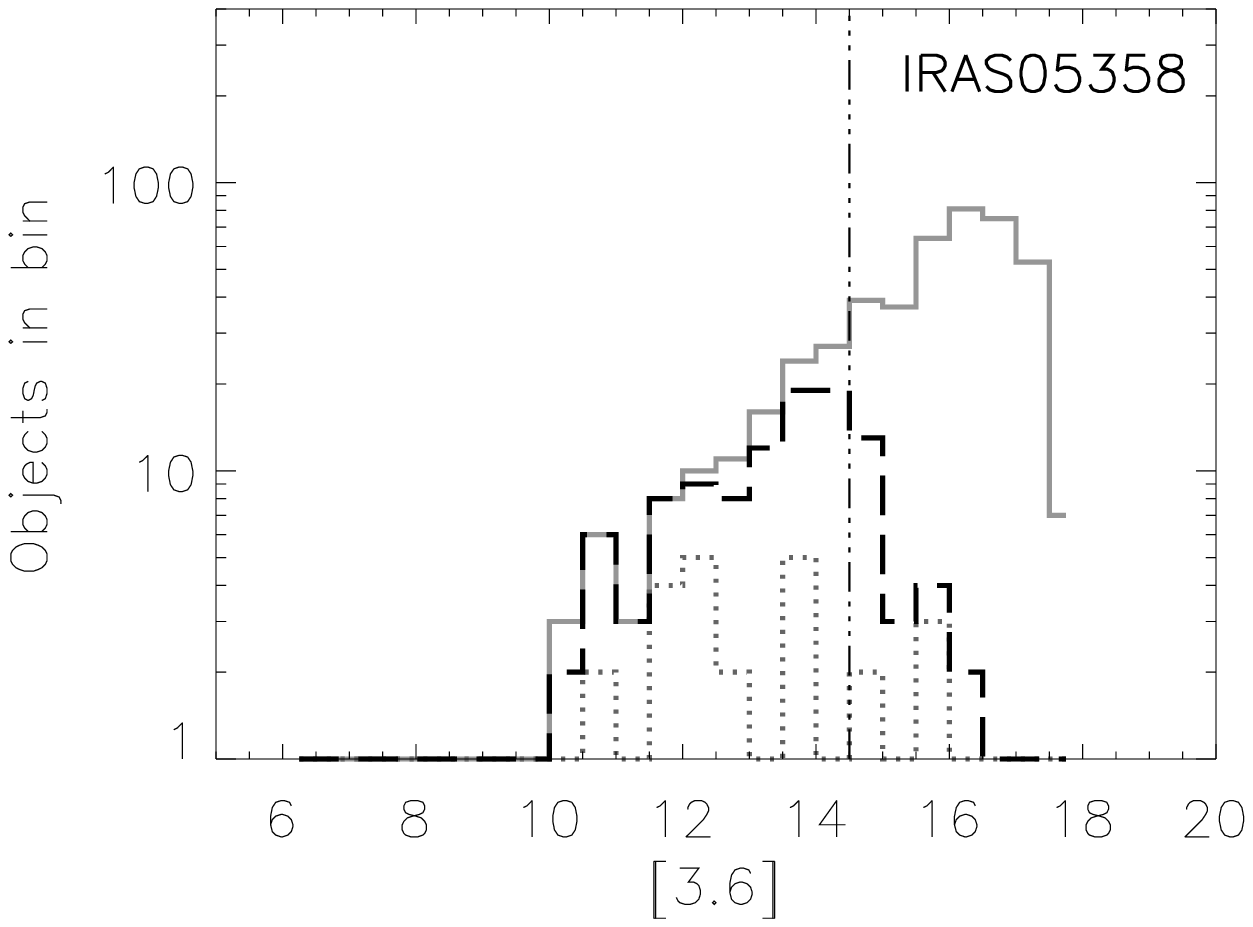}\plotone{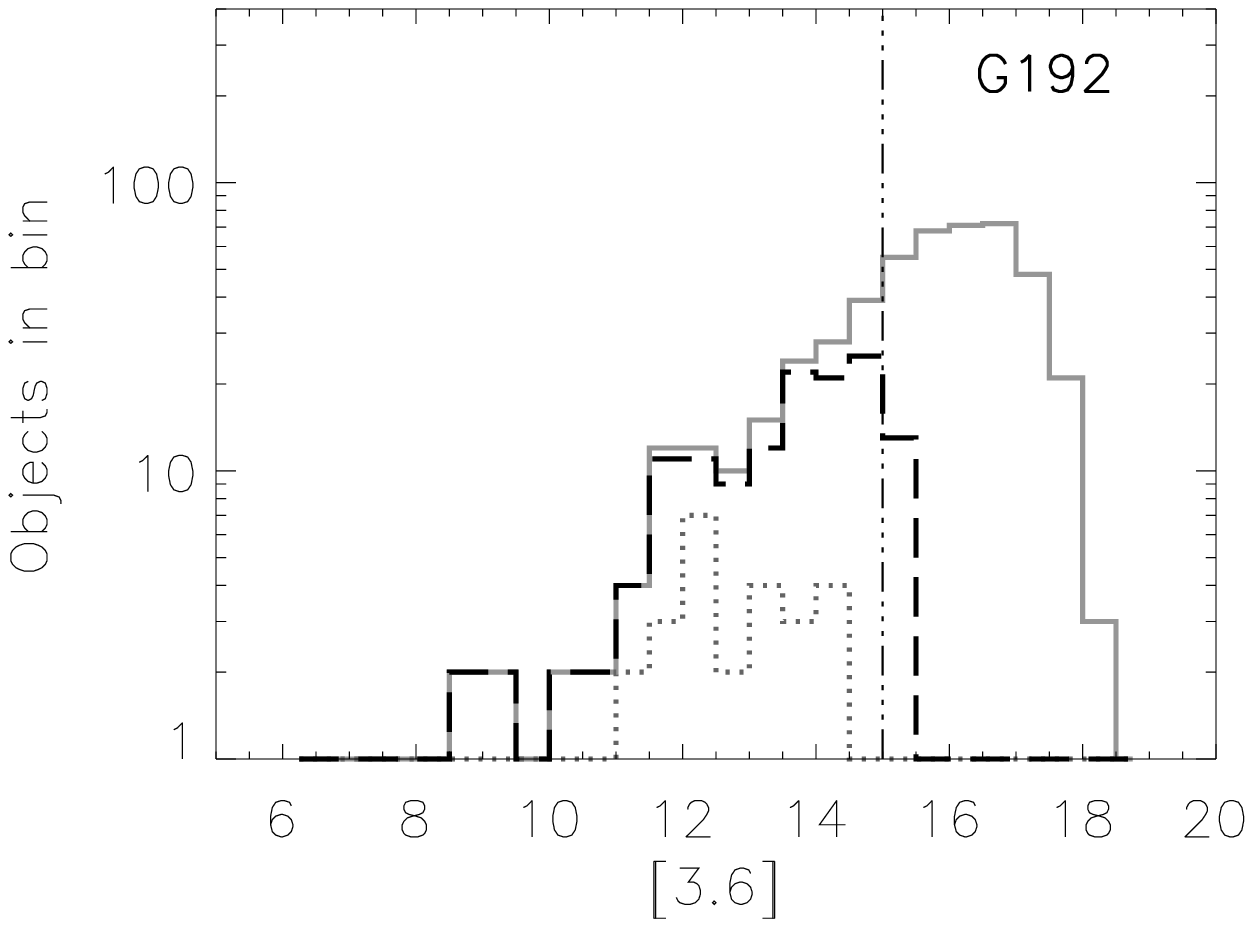}
\plotone{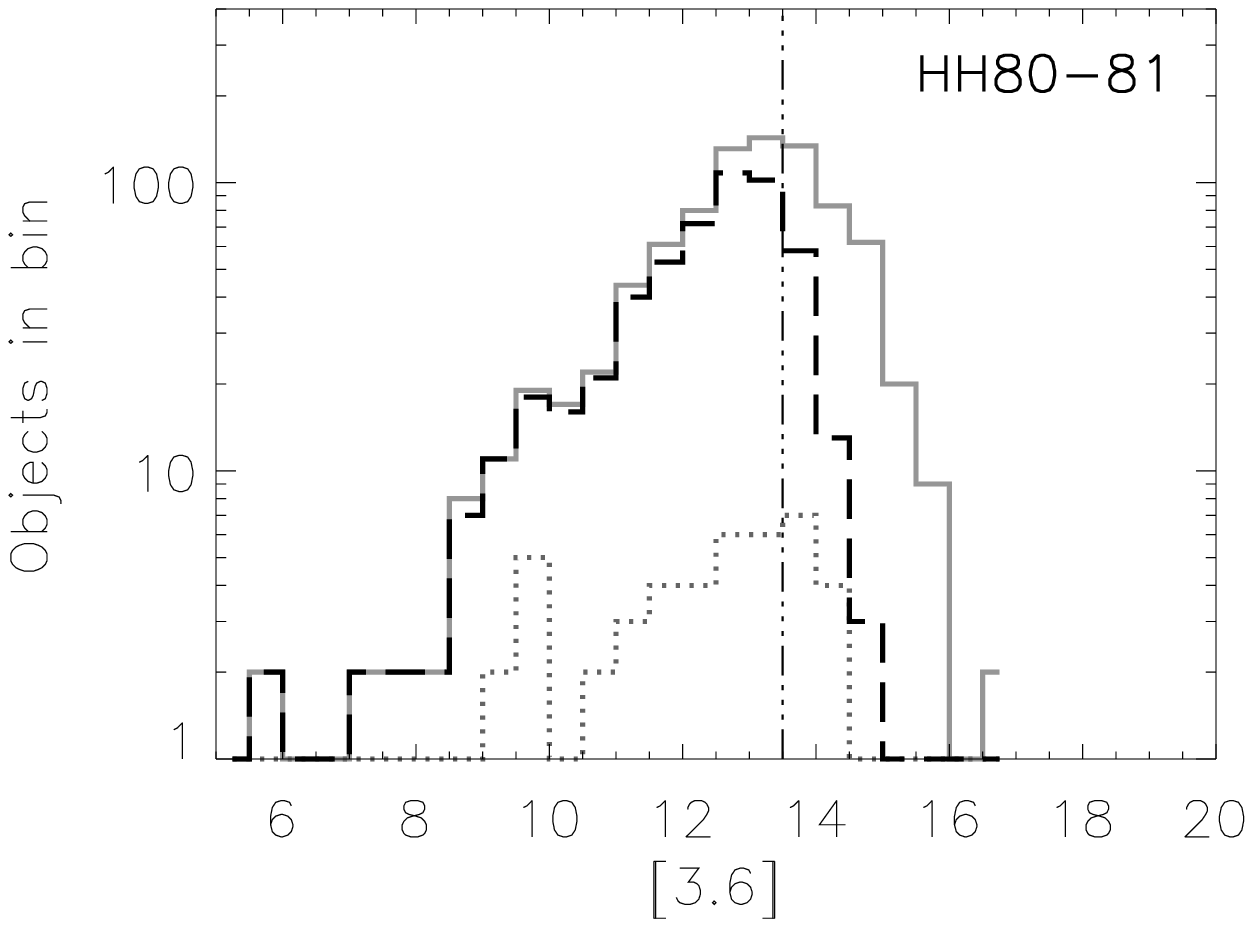}\plotone{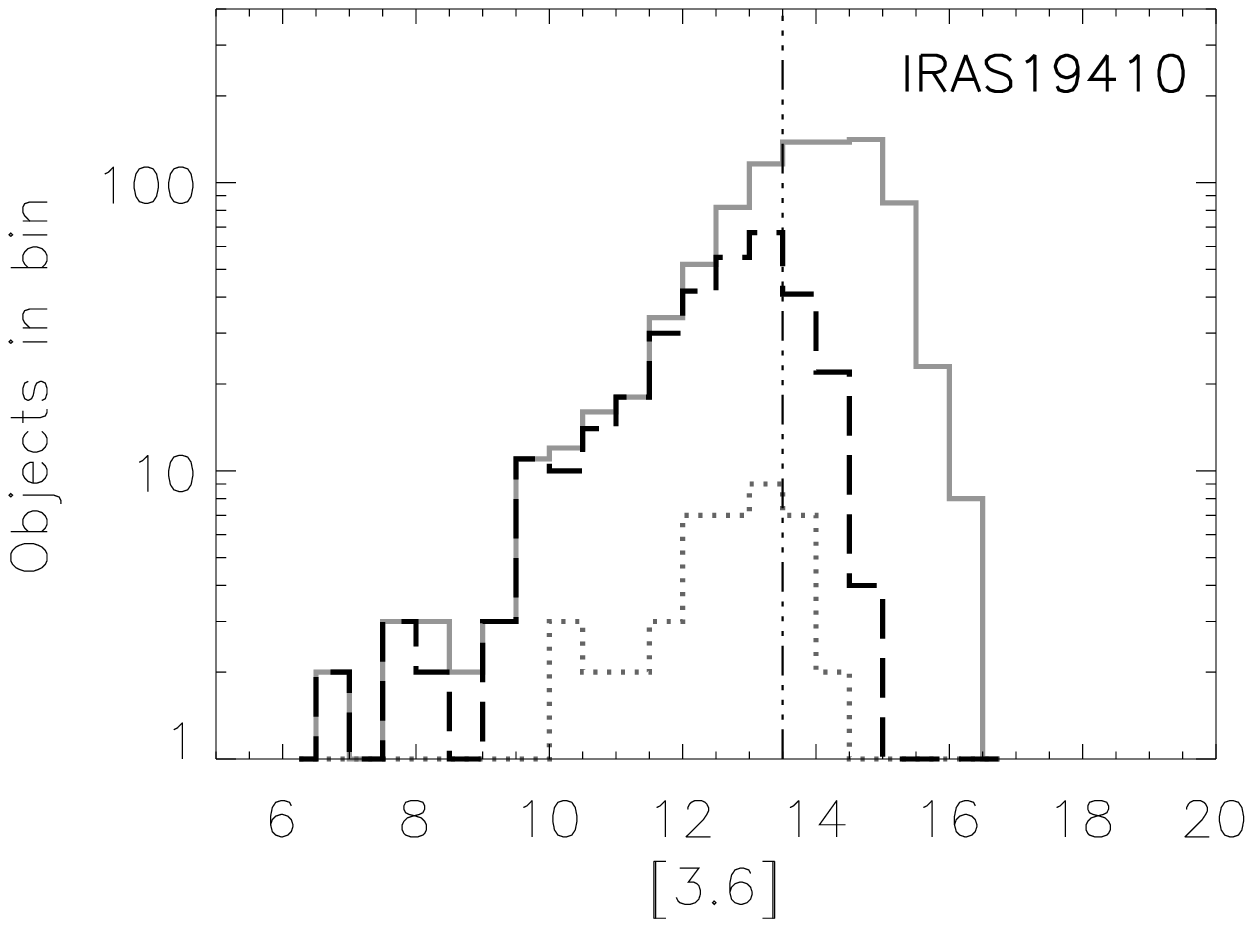}\plotone{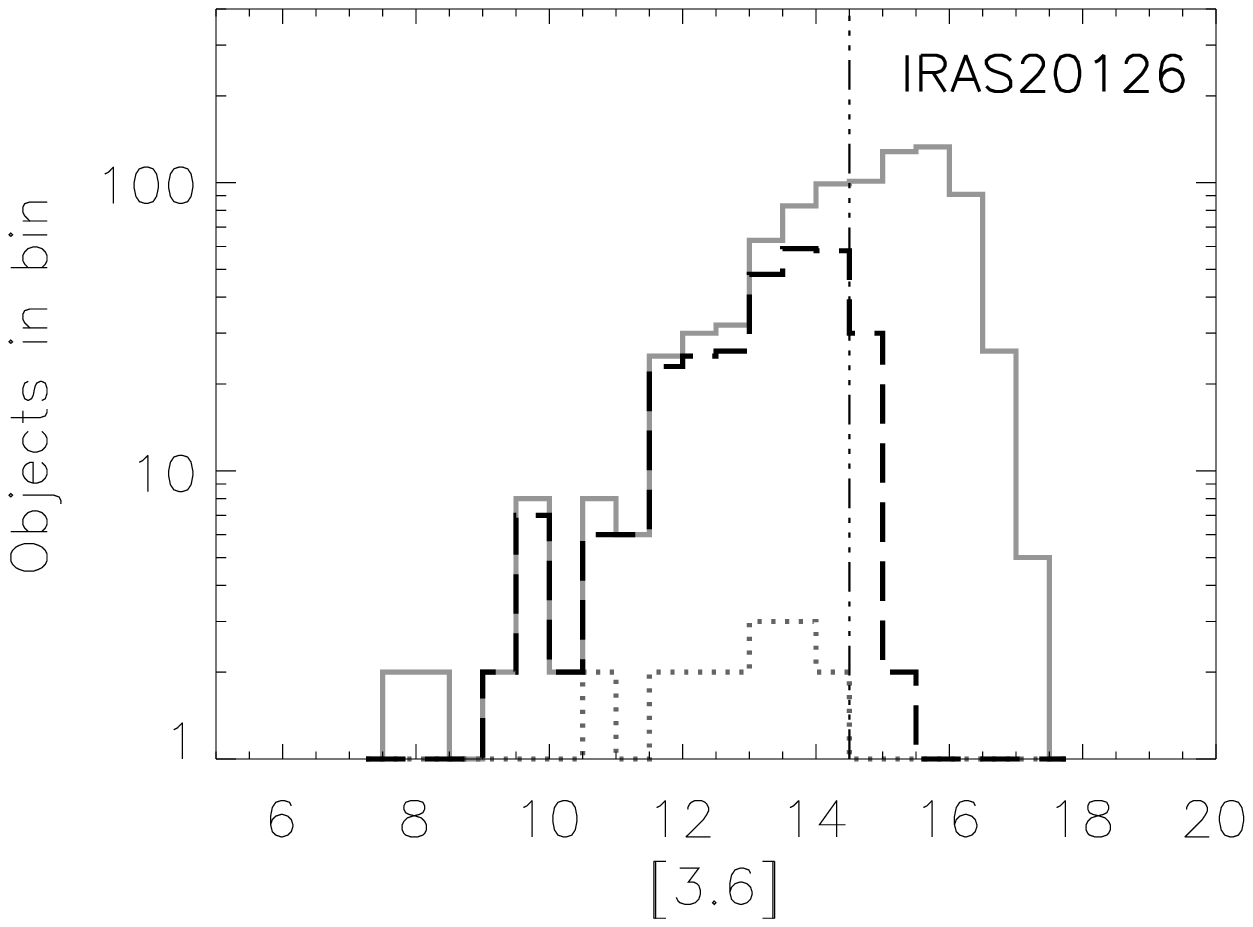}
\plotone{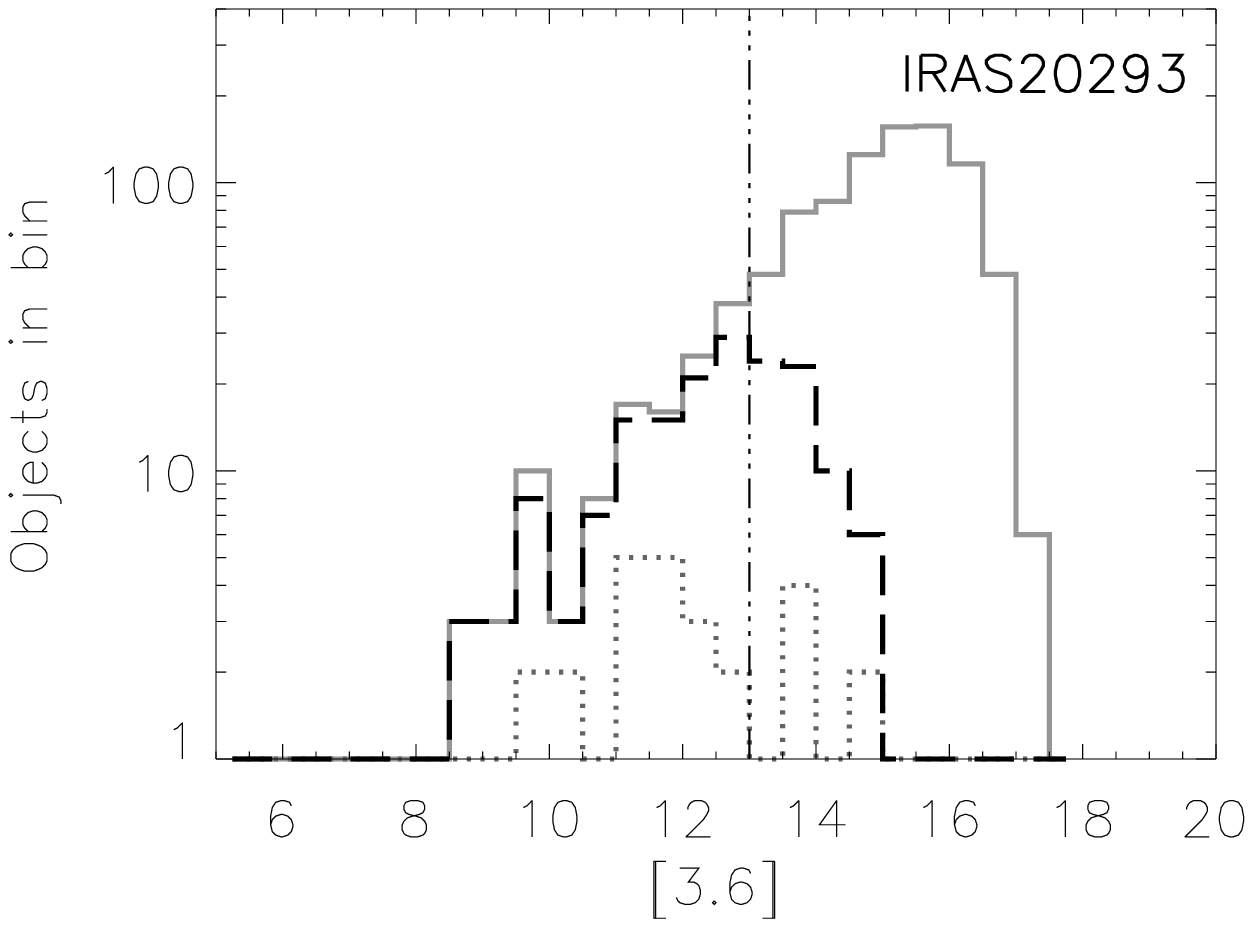}\plotone{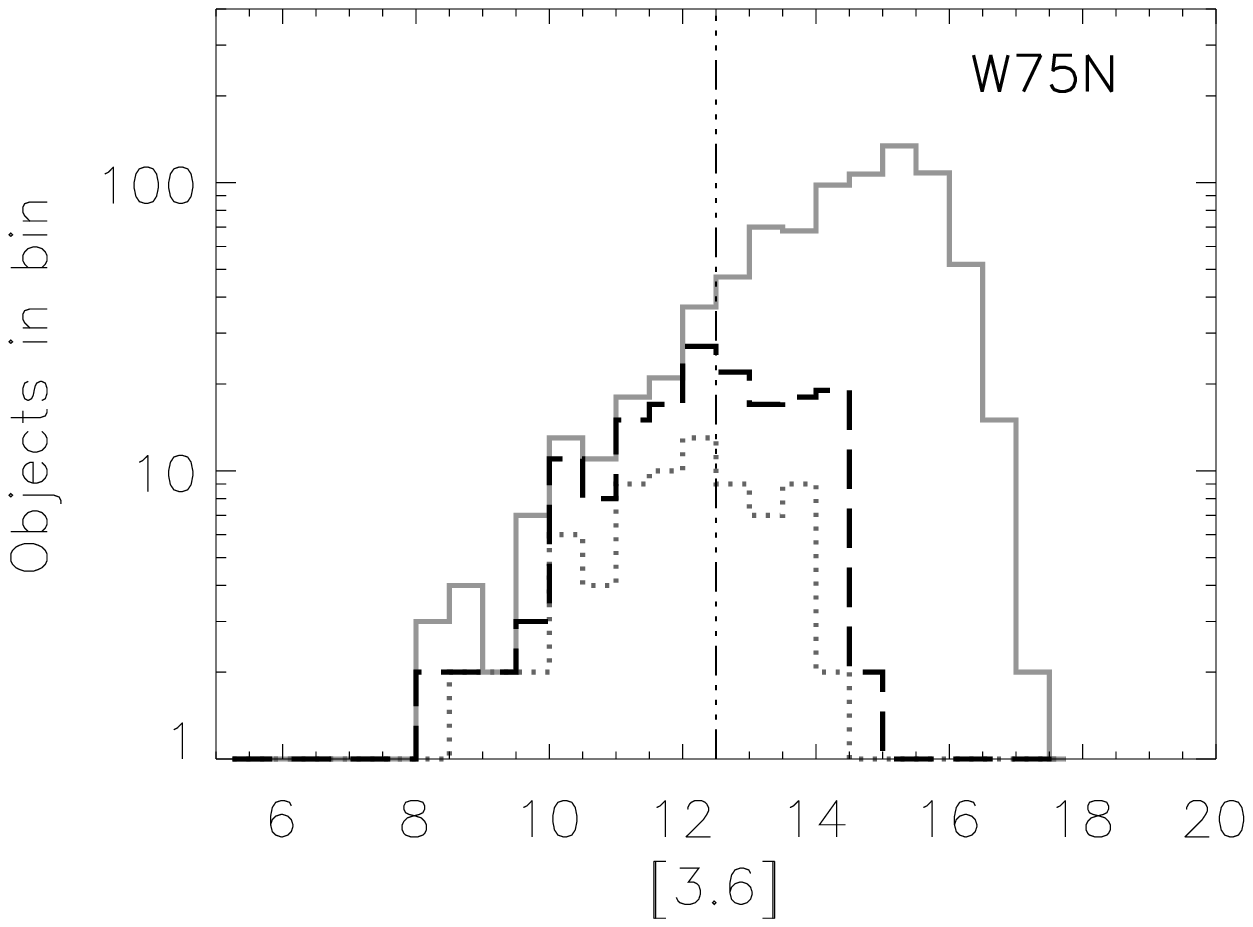}\plotone{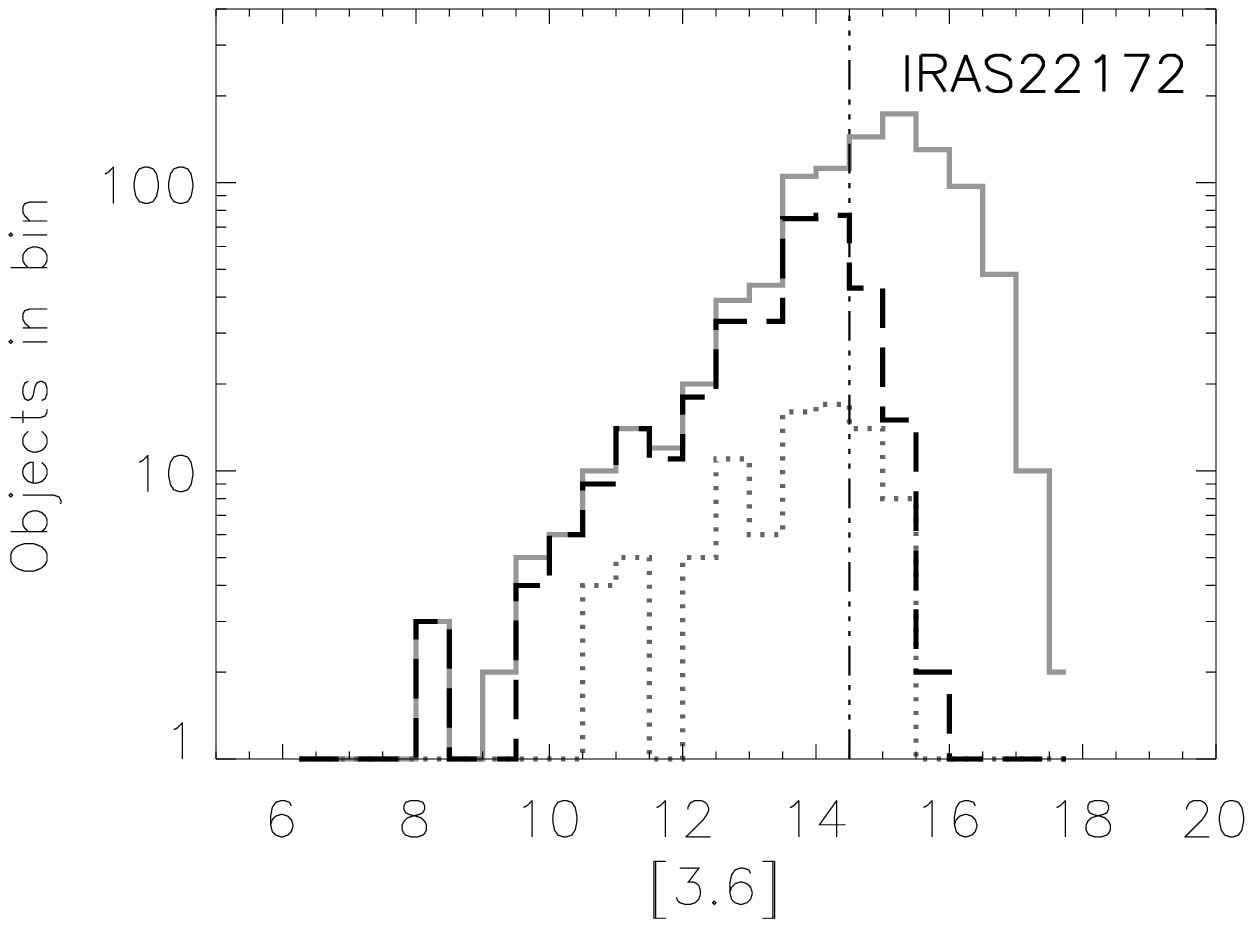}
\caption{Histograms of sources in the 0.5 mag bins as a function
of the 3.6 $\mu$m magnitudes. The gray solid lines plot all the
3.6 $\mu$m detections in each field. The dark dashed lines show
sources with multi-band photometry, which can be placed on one or
more color-color diagrams for search of infrared excess. The gray
dotted lines represent sources that are identified as YSOs.
\label{complete}}
\end{figure}

\clearpage

\begin{figure}
\epsscale{.6} \plotone{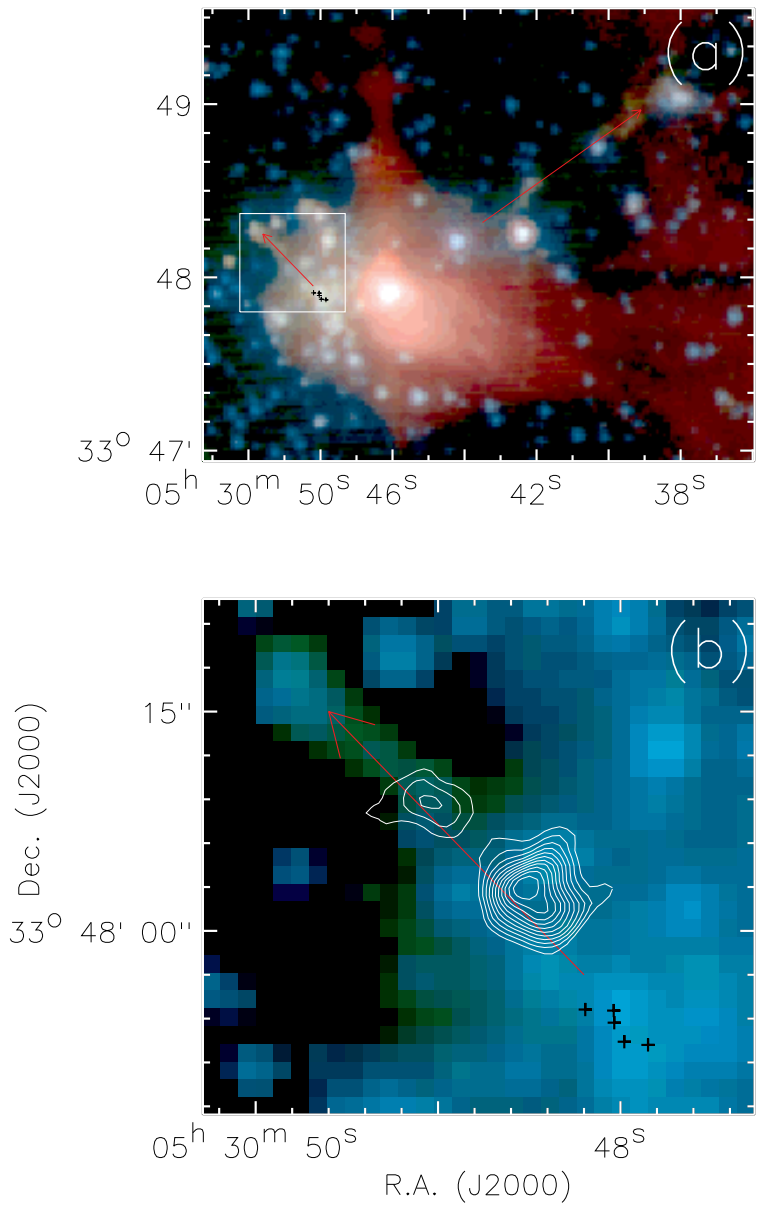} \caption{AFGL 5142: (a) The
3.6/4.5/8.0 $\mu$m (B/G/R) three-color composite of the central
part. The short and long arrows mark the orientations of two
detected jets. (b) The 3.6/4.5 $\mu$m (B/G) two-color composite of
the area labelled by a rectangle in (a), to more clearly show the
short jet prominent in the 4.5 $\mu$m band. Contours are
high-velocity CO emission from outflow B in \citet{Zhang07}. Plus
signs in both panels mark the millimeter continuum peaks (MM1 to
MM5) from \citet{Zhang07}. \label{afgl5142}}
\end{figure}

\clearpage

\begin{figure}
\plotone{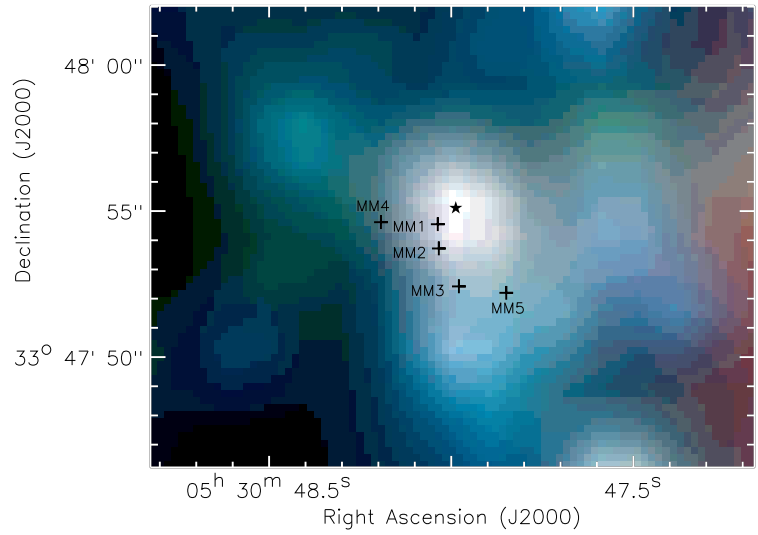} \caption{AFGL 5142: the 3.6/4.5/8.0 $\mu$m
(B/G/R) three-color composite of the inner massive star formation
site, to show a comparison between the point-source detections in
the IRAC bands and previous millimeter continuum observation. The
star symbol denotes a protostar identified from the color-color
diagrams. Plus signs mark the millimeter continuum peaks (MM1 to
MM5) from \citet{Zhang07}. \label{5142core}}
\end{figure}

\clearpage

\begin{figure}
\plotone{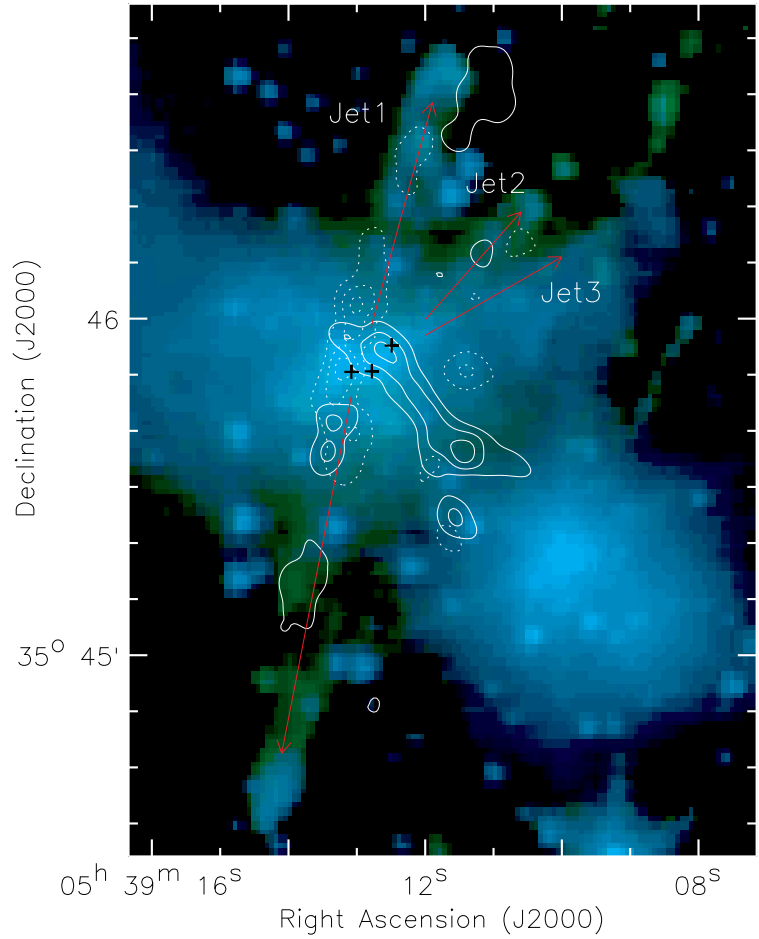} \caption{IRAS 05358: the 3.6/4.5 $\mu$m two-color
composite of the central part. The arrows mark the orientations of
three detected jets, which are labelled as ``Jet1'', ``Jet2'', and
``Jet3'' from east to west. Solid and dashed contours show blue-
and redshifted lobes of CO outflows from \citet{Beuther02},
repectively. Plus signs mark the 3.1 mm continuum peaks from
\citet{Beuther07}. \label{i05358}}
\end{figure}

\clearpage

\begin{figure}
\epsscale{1.} \plotone{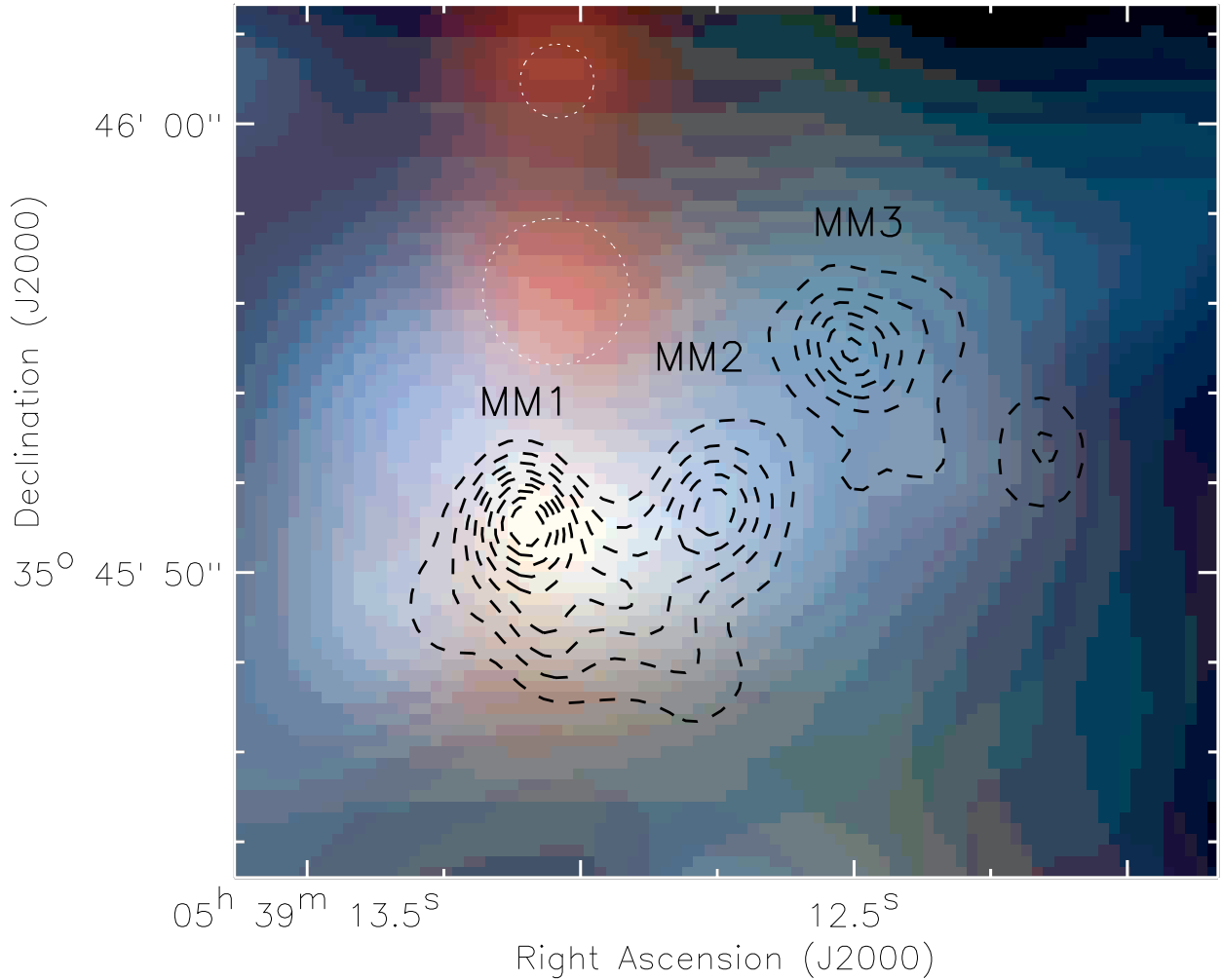} \caption{IRAS 05358: the
3.6/4.5/8.0 $\mu$m three-color composite of the inner massive star
formation site, to show a comparison between the point-source
detections in the IRAC bands and previous millimeter continuum
observation. The dashed contours are for the 3.1 mm continuum
emission with three peaks labelled as ``MM1'', ``MM2'', ``MM3''
\citep{Beuther07}. The two white dashed circles denote
``bandwidth'' artifacts in the 8.0 $\mu$m band. \label{05358core}}
\end{figure}

\clearpage

\begin{figure}
\plotone{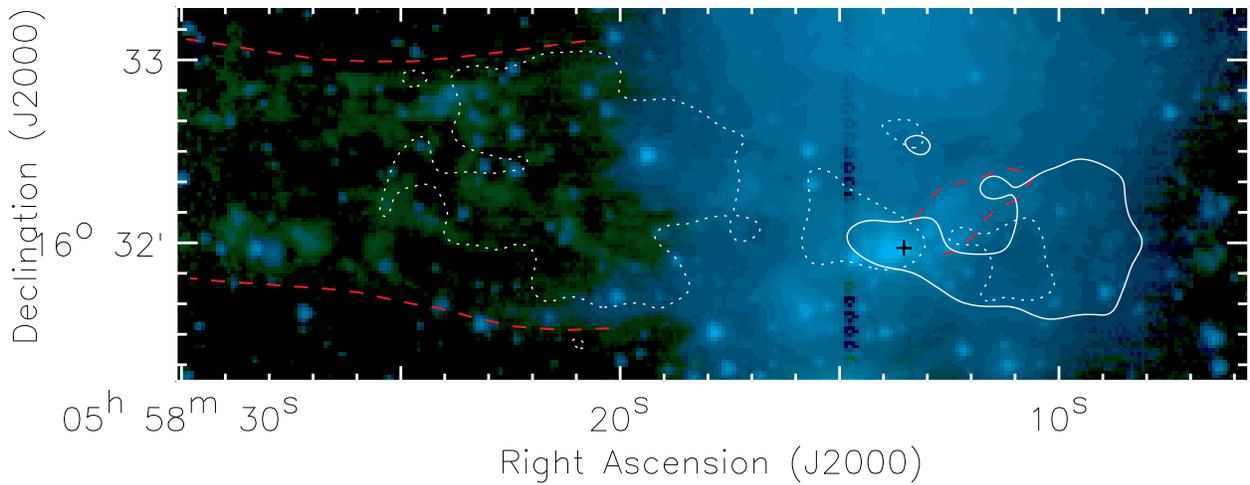} \caption{G192: the 3.6/4.5 $\mu$m two-color
composite of the central part. Solid and dashed contours are the
lowest contours of red- and blueshifted CO outflow, respectively
\citep{Shepherd98}. Two red dashed lines in the east outline the
area filled with ``green'' nebulosities, and the red dashed curve
in the west the candle-flame-shaped structure (this structure is
more prominent in the three-color composite Figure \ref{fig2}c).
The plus sign marks the millimeter continuum peak from
\citet{Shepherd98}. \label{g192}}
\end{figure}

\clearpage

\begin{figure}
\epsscale{1.} \plotone{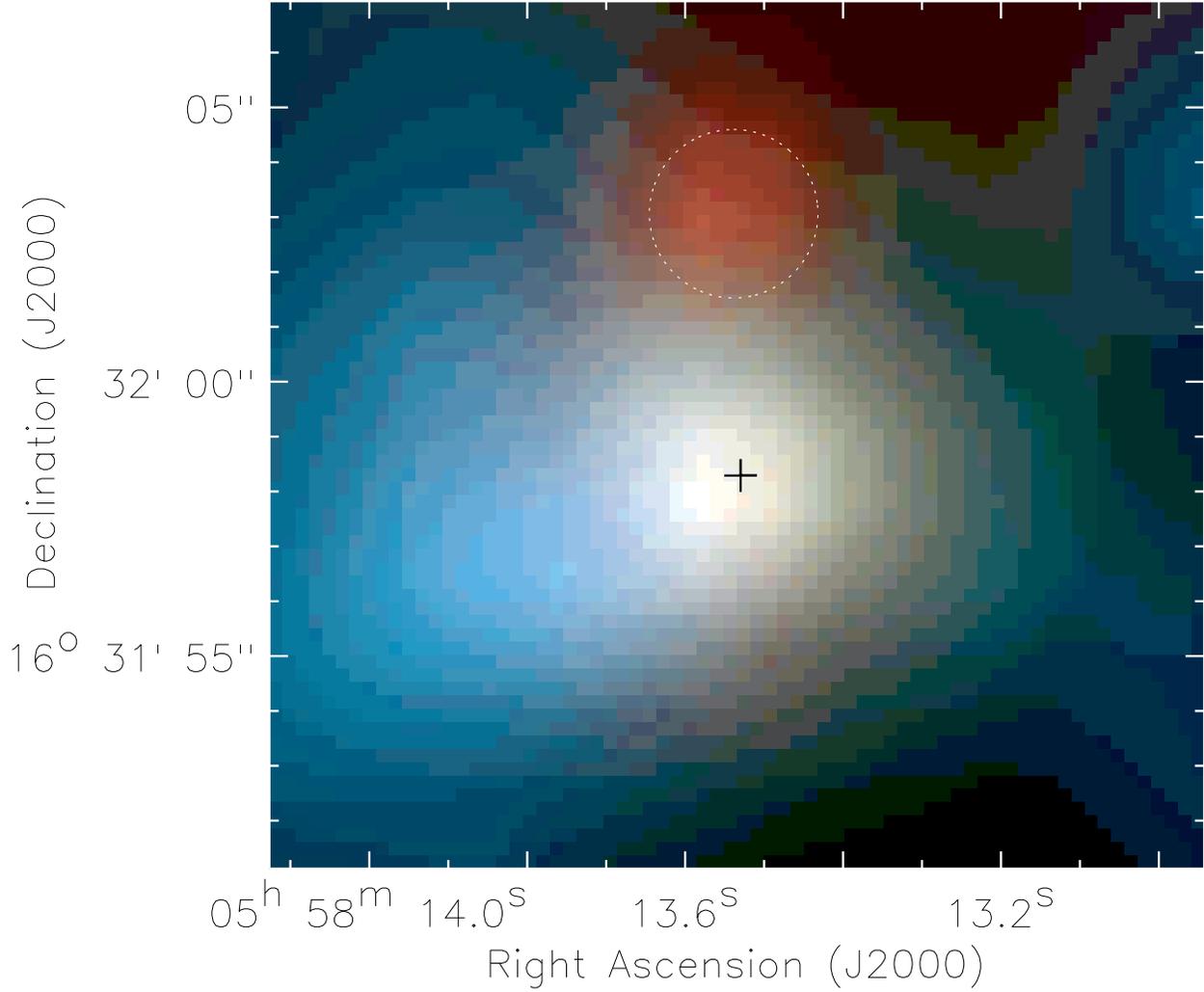} \caption{G192: the 3.6/4.5/8.0
$\mu$m three-color composite of the inner massive star formation
site, to show a comparison between the point-source detections in
the IRAC bands and previous millimeter continuum observation. The
plus sign marks the millimeter continuum peak from
\citet{Shepherd98}. \label{192core}}
\end{figure}

\clearpage

\begin{figure}
\epsscale{.8} \plotone{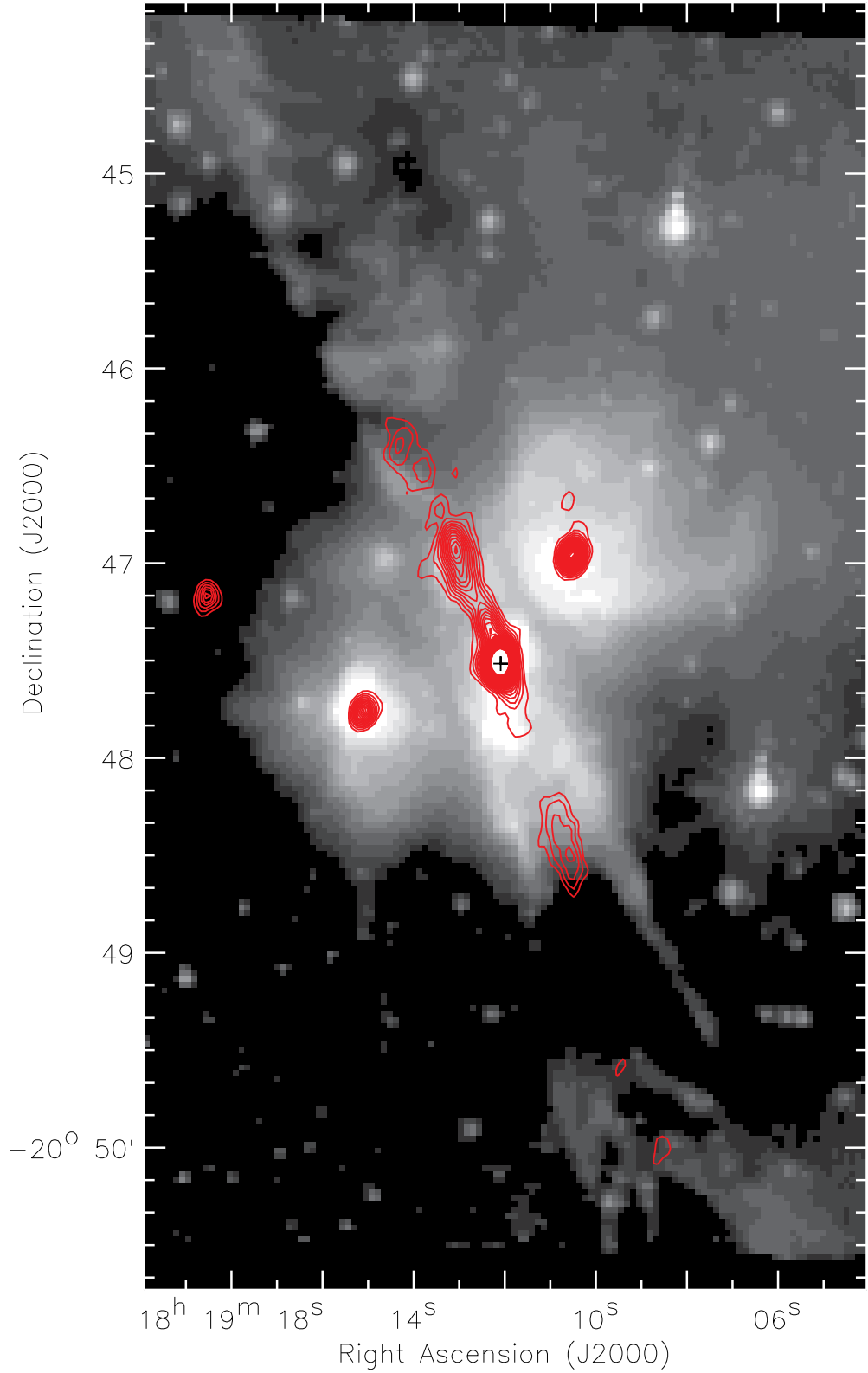} \caption{HH 80-81: the 8.0 $\mu$m
band emission in grayscale and the 6 cm continuum emission from
\citet{Marti93} in contours. The plus sign marks the 7 mm
continuum peak from \citet{Gomez03}. \label{h8081}}
\end{figure}

\clearpage

\begin{figure}
\epsscale{1.} \plotone{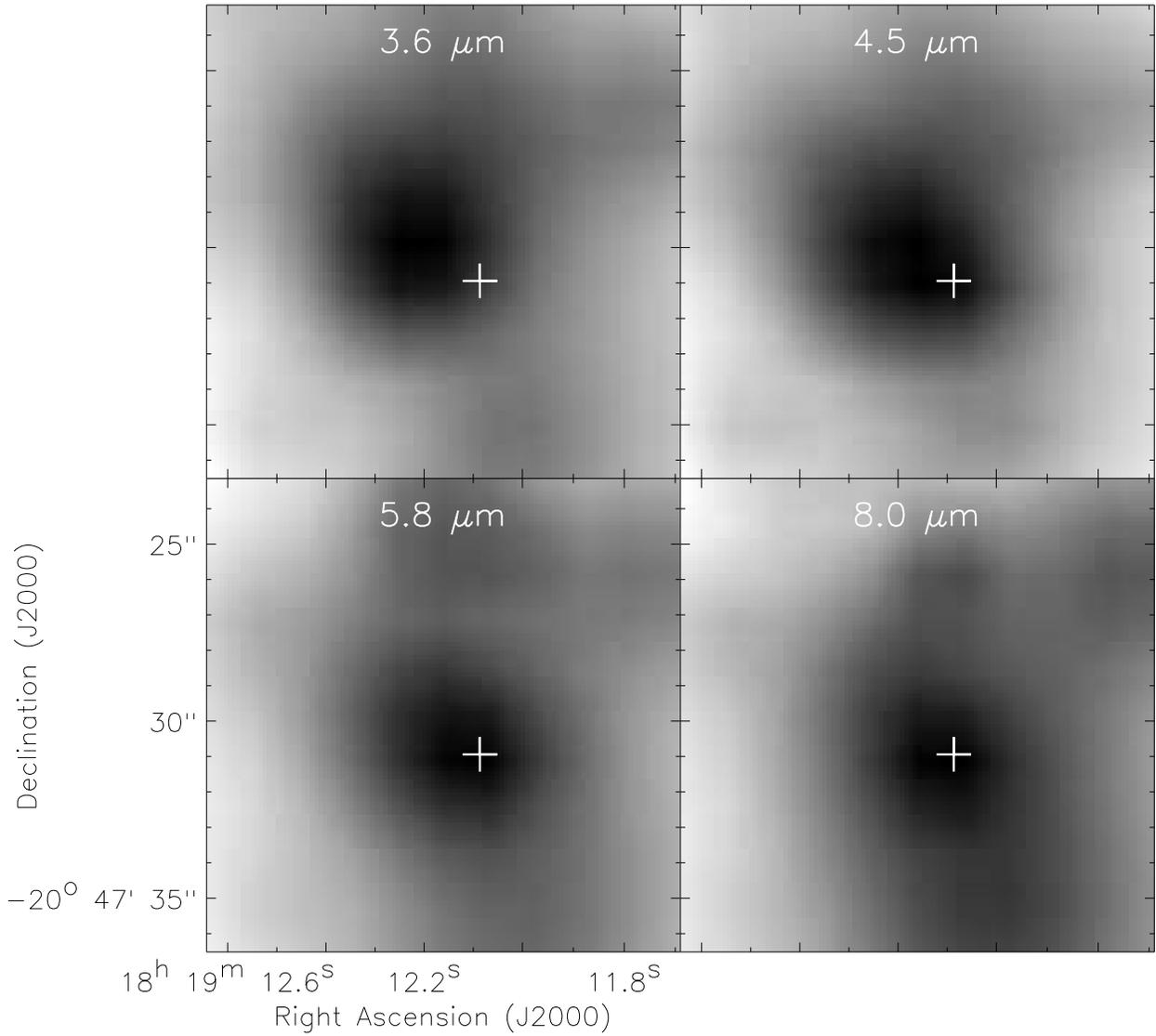} \caption{HH 80-81: the IRAC band
emission from the inner massive star formation site, to show a
comparison between the point-source detections in the IRAC bands
and previous millimeter continuum observation. The plus sign marks
the 7 mm continuum peak from \citet{Gomez03}. \label{8081core}}
\end{figure}

\clearpage

\begin{figure}
\plotone{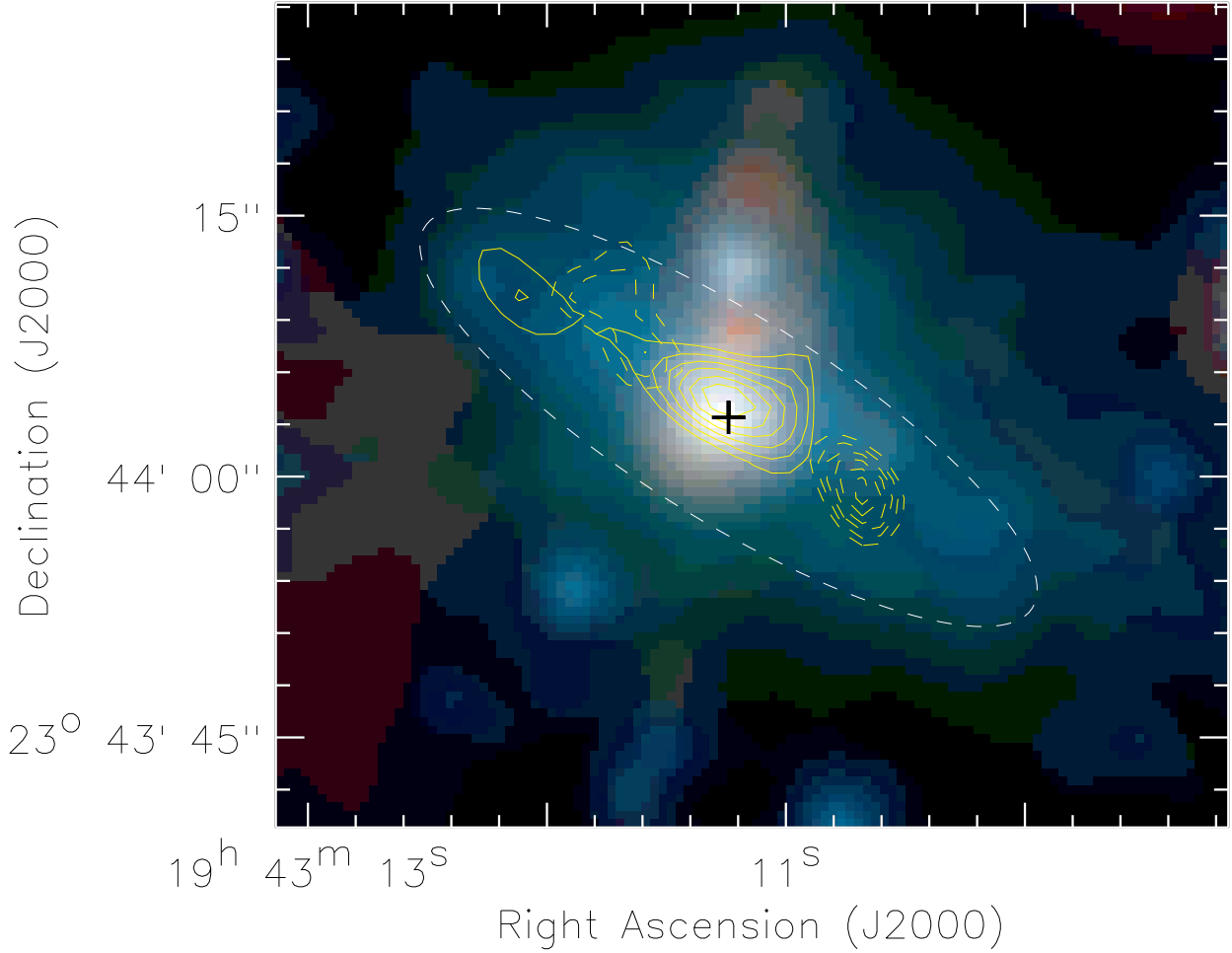} \caption{IRAS 19410: the 3.6/4.5/8.0 $\mu$m
three-color composite of the central part. The dashed ellipse
outlines the IRAC elliptical structure. Solid and dashed contours
show blue- and redshifted lobes from a CO outflow by
\citet{Beuther03}, repectively,  and the plus sign a millimeter
continuum peak therein. \label{i19410}}
\end{figure}

\clearpage

\begin{figure}
\plotone{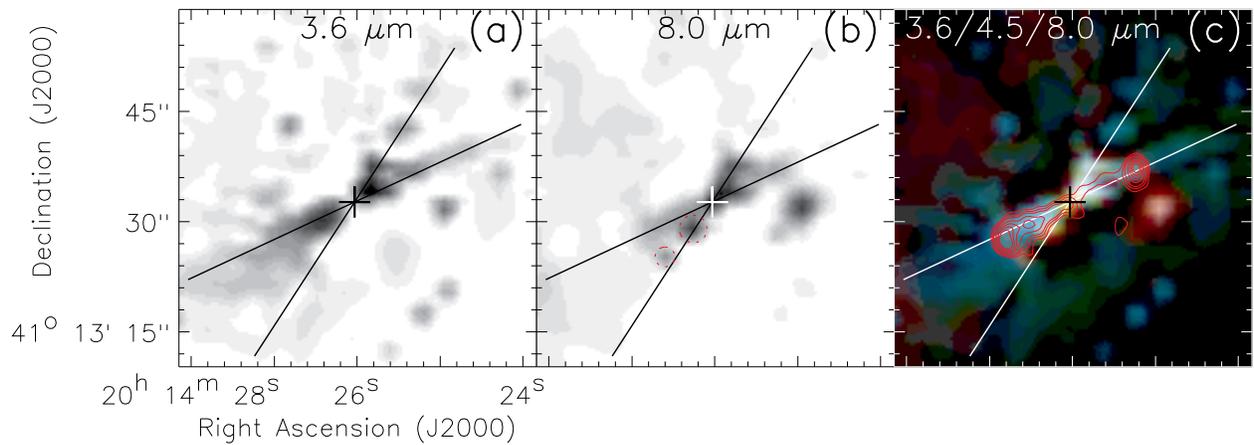} \caption{IRAS 20126: the IRAC bands emission
from the central part. (a) The 3.6 $\mu$m emission. To highlight
the bipolar nebula a high-pass-filtering in IDL has been applied
to the image. (b) Same as (a), but for the 8.0 $\mu$m emission.
Infrared emission from the central source is detected at this
wavelength. The two dashed circles mark ``bandwidth'' artifacts
induced by the central bright source. (c) The 3.6/4.5/8.0 $\mu$m
three-color composite, with an SiO (2-1) jet from
\citet{Cesaroni99} overlaid in red contours. In each panel the
crossing lines outline the wall of the IRAC biconical structure.
The plus sign marks the 1.2 mm continuum peak from
\citet{Cesaroni05}. \label{i20126}}
\end{figure}

\clearpage

\begin{figure}
\epsscale{1.} \plotone{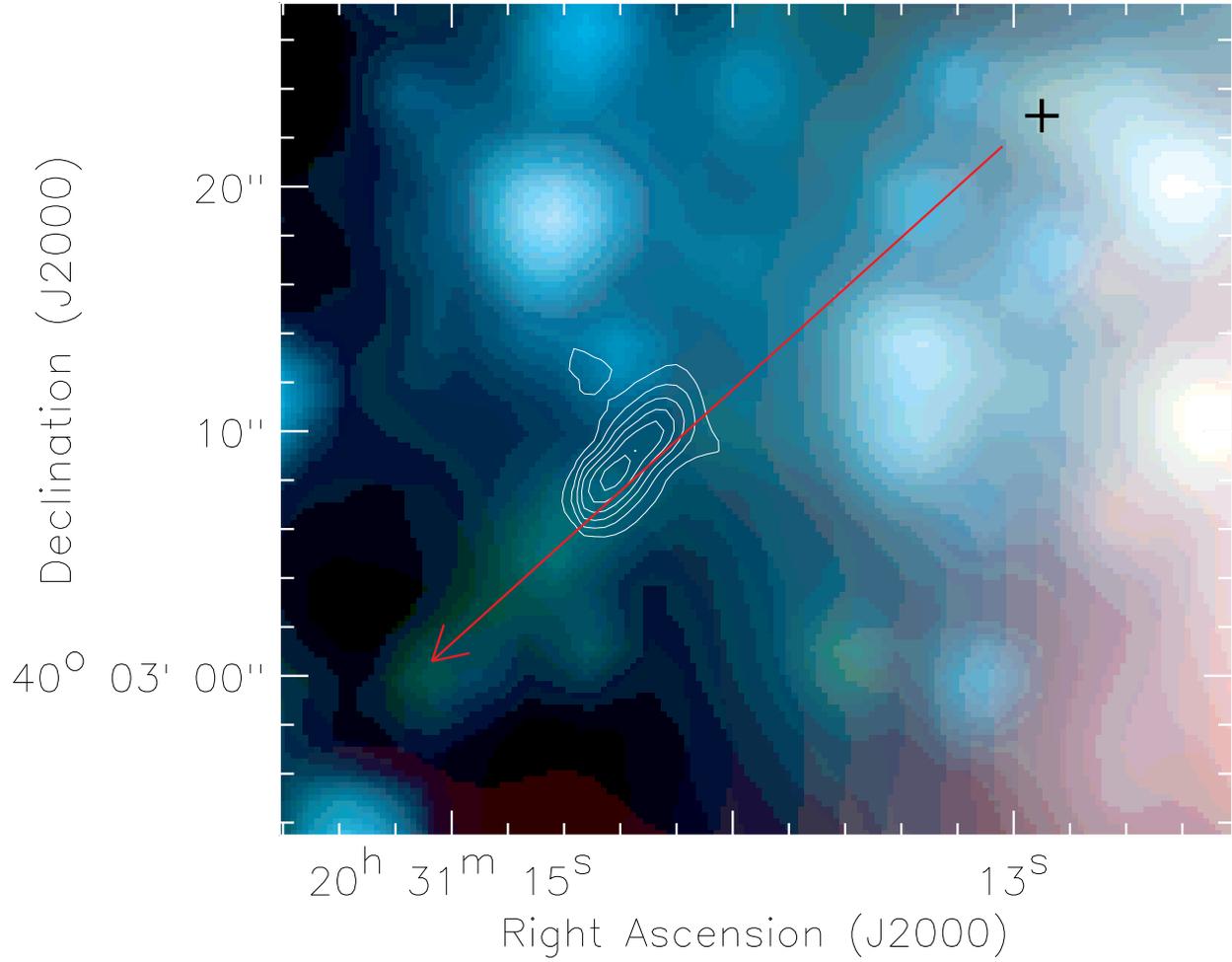} \caption{IRAS 20293: the
3.6/4.5/8.0 $\mu$m three-color composite of the central part. The
arrow marks the orientation of a jet prominent in the 4.5 $\mu$m
band. The contours show blueshifted high-velocity CO emission in a
a jet-like outflow and the plus sign the millimeter continuum peak
from \citet{Beuther04a}. \label{i20293}}
\end{figure}

\clearpage

\begin{figure}
\epsscale{1.} \plotone{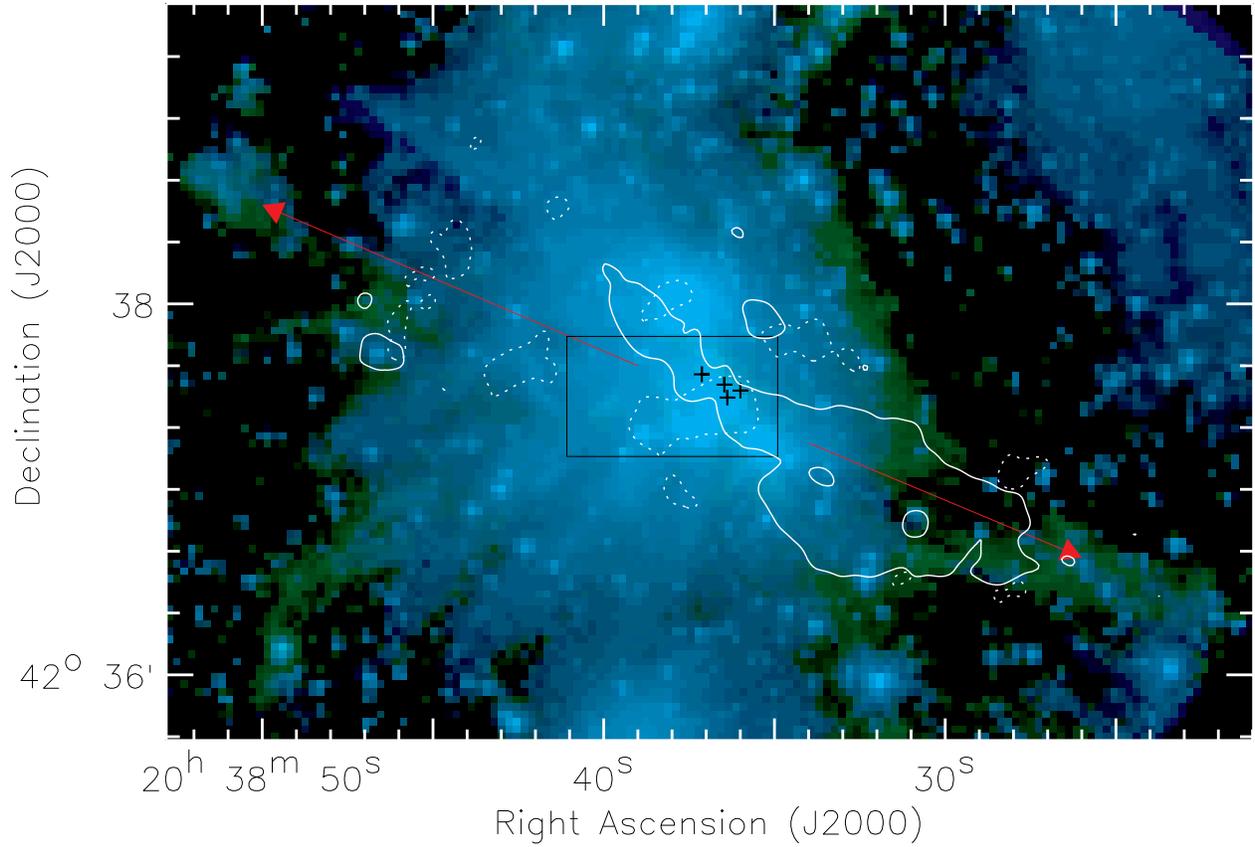} \caption{W75\,N: the 3.6/4.5
$\mu$m two-color composite of the central part. The double arrows
mark the orientation of a large scale bow-shaped structure. Solid
and dashed contours are the lowest contours of red- and
blueshifted emission in a CO outflow, respectively
\citep{Shepherd03}. Plus signs mark the millimeter continuum peaks
from \citet{Shepherd01}.\label{w75}}
\end{figure}

\clearpage

\begin{figure}
\epsscale{.6} \plotone{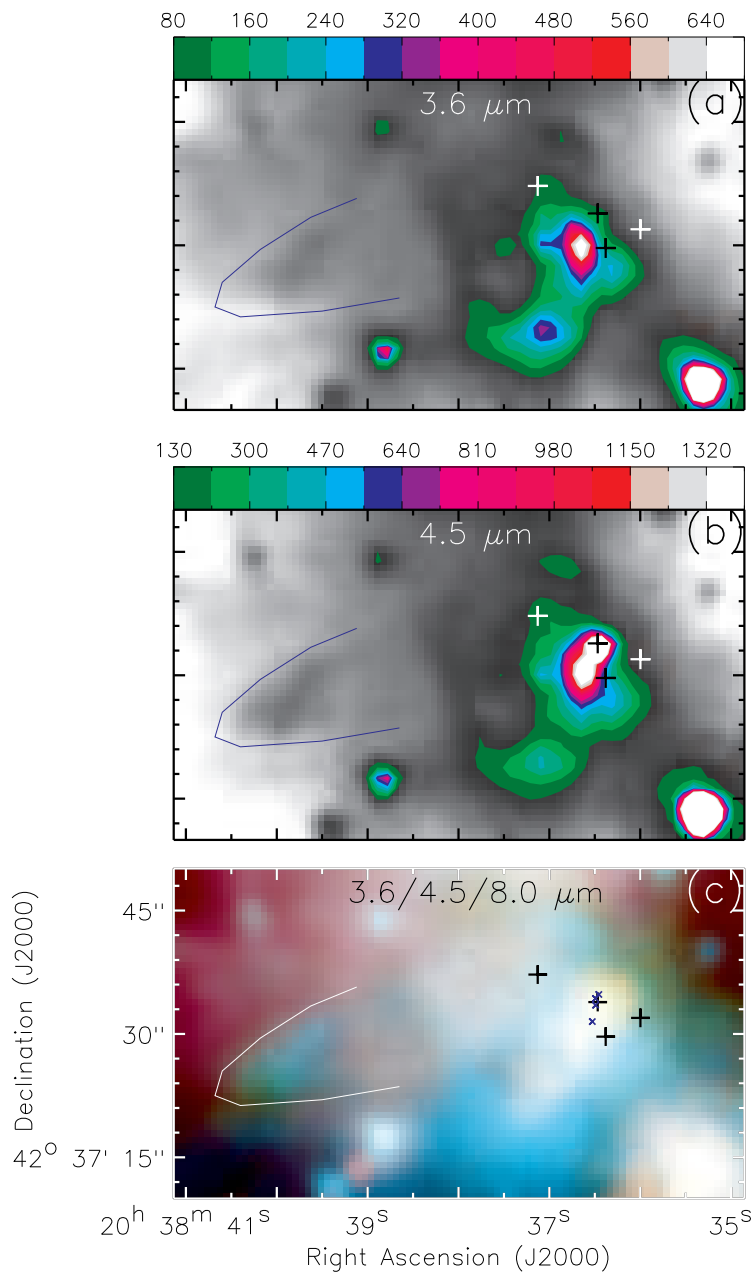} \caption{W75\,N: the IRAC bands
emission of the area labelled as a rectangle in Figure \ref{w75}.
(a) The 3.6 $\mu$m emission, with emission levels between 8.2 and
80 MJy/sr shown in grayscale and that beyond 80 MJy/sr in colored
contours. The contour levels in MJy/sr are labelled at the top of
the image. (b) Same as (a), but for the 4.5 $\mu$m emission. The
emission between 10.7 and 130 MJy/sr is shown in grayscale and
that beyond 130 MJy/sr in colored contours. (c) The 3.6/4.5/8.0
$\mu$m three-color composite. In each panel a solid curve outlines
the edge of a small bow-shaped structure and the plus signs mark
the millimeter continuum peaks (MM1 to MM4) from
\citet{Shepherd01}. The crosses in (c) mark the centimeter
continuum peaks revealed by \citet{Hunter95} and
\citet{Torrelles97}. \label{w75n}}
\end{figure}

\clearpage

\begin{figure}
\epsscale{.8} \plotone{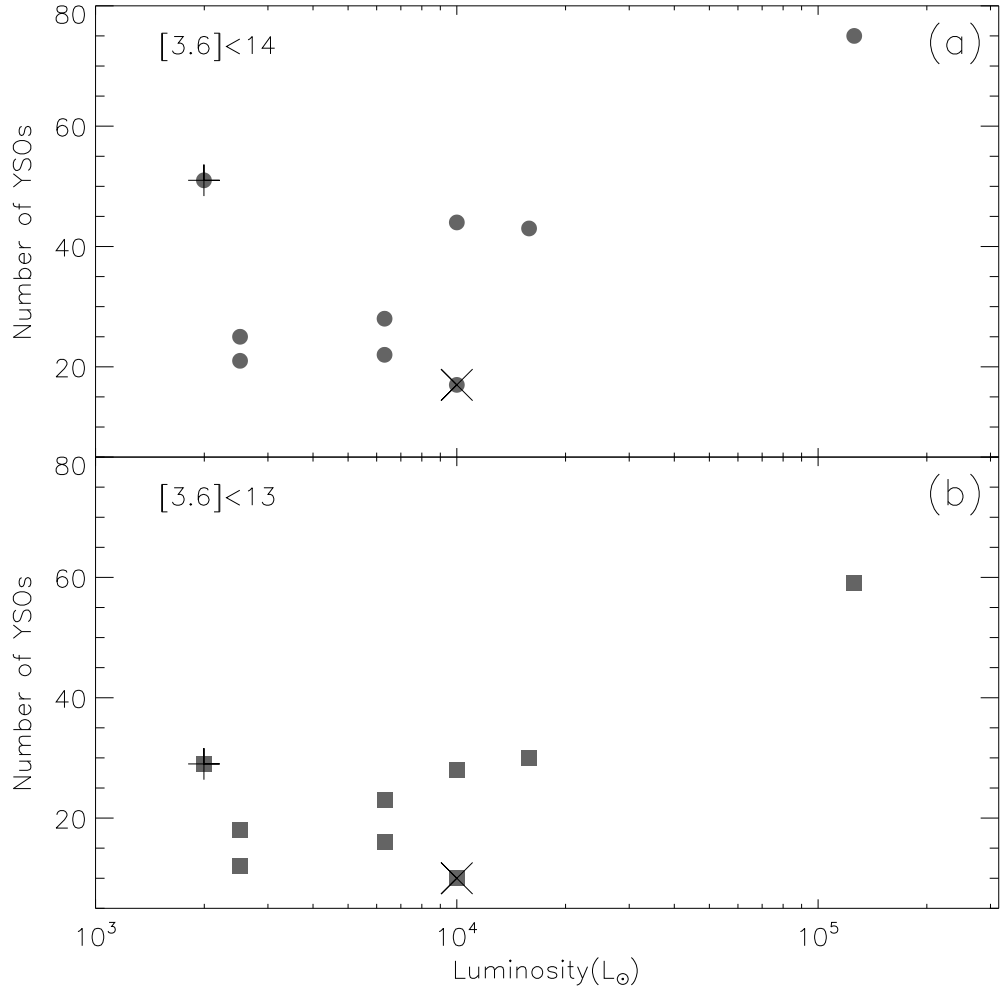} \caption{The number of detected
YSOs vs. the IRAS luminosity. (a) The YSOs number counted from
those with $[3.6]<14$. The data points for IRAS 22172 and IRAS
20126 are overlaid with a plus sign and a cross, respectively.
(b)Same as (a), but for YSOs with $[3.6]<13$. \label{fig17}}
\end{figure}

\end{document}